\documentclass[aps,pra,preprint,amsmath,amssymb,longbibliography]{revtex4-1}
\usepackage{graphicx,epsfig,epsf,color,hhline,import}
\usepackage{dcolumn}
\usepackage{bm}
\usepackage{tensor,pbox}
\usepackage{braket}
\usepackage{tikz}
\usepackage{epstopdf}
\usepackage{hyperref}
\usepackage[a4paper,
width=185mm,
top=15mm,bottom=15mm,
includehead, 
includefoot,
bindingoffset=6mm]{geometry}

\usepackage{changes}
\usepackage{cancel}
\usepackage{color}

\newcommand{\bea}{\begin{eqnarray}}
\newcommand{\ea}{\end{eqnarray}}
\newcommand{\eea}{\end{eqnarray}}

\newcommand{\sumint}[1]

\newcommand{\bOne}{\mbox{\bf 1}}
 
\begin{document}
\newcommand{\rem}[1]{\textcolor{red}{\sout{#1}}}
\newcommand{\add}[1]{\textcolor{blue}{\uline{#1}}}
\newcommand{\ri}{ i}
\newcommand{\re}{ e}
\newcommand{\bx}{{\bm x}}
\newcommand{\bd}{{\bm d}}
\newcommand{\be}{{\bm e}}
\newcommand{\br}{{\bm r}}
\newcommand{\bk}{{\bm k}}
\newcommand{\bA}{{\bm A}}
\newcommand{\bD}{{\bm D}}
\newcommand{\bE}{{\bm E}}
\newcommand{\bB}{{\bm B}}
\newcommand{\bI}{{\bm I}}
\newcommand{\bH}{{\bm H}}
\newcommand{\bL}{{\bm L}}
\newcommand{\bR}{{\bm R}}
\newcommand{\bZero}{{\bm 0}}
\newcommand{\bM}{{\bm M}}
\newcommand{\bX}{{\bm X}}
\newcommand{\bn}{{\bm n}}
\newcommand{\bs}{{\bm s}}
\newcommand{\bv}{{\bm v}}
\newcommand{\tbs}{\tilde{\bm s}}
\newcommand{\rSi}{{\rm Si}}
\newcommand{\beps}{\mbox{\boldmath{$\epsilon$}}}
\newcommand{\bGamma}{\mbox{\boldmath{$\Gamma$}}}
\newcommand{\bxi}{\mbox{\boldmath{$\xi$}}}
\newcommand{\rg}{{\rm g}}
\newcommand{\tr}{{\rm tr}}
\newcommand{\xmax}{x_{\rm max}}
\newcommand{\xb}{\overline{x}}
\newcommand{\pb}{\overline{p}}
\newcommand{\ra}{{\rm a}}
\newcommand{\rx}{{\rm x}}
\newcommand{\rs}{{\rm s}}
\newcommand{\rP}{{\rm P}}
\newcommand{\up}{\uparrow}
\newcommand{\down}{\downarrow}
\newcommand{\hc}{H_{\rm cond}}
\newcommand{\kb}{k_{\rm B}}
\newcommand{\cI}{{\cal I}}
\newcommand{\tit}{\tilde{t}}
\newcommand{\cE}{{\cal E}}
\newcommand{\cC}{{\cal C}}
\newcommand{\Ubs}{U_{\rm BS}}
\newcommand{\sech}{{\rm sech}}
\newcommand{\qq}{{\bf ???}}
\newcommand*{\etal}{\textit{et al.}}
\def\vec#1{\bm{#1}}
\def\ket#1{|#1\rangle}
\def\bra#1{\langle#1|}
\def\keps{\bm{k}\boldsymbol{\varepsilon}}
\def\dm{\boldsymbol{\wp}}
\newcommand{\pscal}[2]{\ensuremath{ \langle \, #1 \, \vert  \, #2 \, \rangle}} 
\newcommand{\dens}[2]{\ensuremath{ \vert \, #1 \, \rangle \langle \, #2 \, \vert}} 
\newcommand{\moy}[1]{\ensuremath{ \langle \, #1 \, \rangle}} 
\newcommand{\moyvec}[2]{\ensuremath{ \langle #2 \, | \, #1 \, | \, #2 \rangle}} 
\newcommand{\com}[2]{\ensuremath{ \left[  #1 , #2  \right] }} 
\newcommand{\acom}[2]{\ensuremath{ \left\{  #1 , #2  \right\} }} 
\newcommand{\paran}[1]{\ensuremath{ \left(  #1  \right) }} 
\newcommand{\jmoins}{$J_{-} \,$}
\newcommand{\jplus}{$J_{+} \,$}
\newcommand{\jx}{$J_{x} \,$}
\newcommand{\jy}{$J_{y} \,$}
\newcommand{\jz}{$J_{z} \,$}
\newcommand{\lu}{\textcolor{blue}}
\newcommand{\lug}{\textcolor{green}}
\renewcommand{\figurename}{Figure}

\makeatletter
\def\blfootnote{\gdef\@thefnmark{}\@footnotetext}
\makeatother

\title{Quantum metrology with quantum-chaotic sensors}
\author{Lukas J.~Fiderer$^{1}$ $\&$ Daniel Braun$^{1}$}
\affiliation{$^1$Institute for Theoretical Physics, University of T{\"u}bingen, Auf der Morgenstelle 14, 72076 T{\"u}bingen, Germany\\Correspondence should be addressed to L.J.F. (email:~lukas.fiderer@uni-tuebingen.de) or D.B.~(email:~daniel.braun@uni-tuebingen.de).}

\begin{abstract}
Quantum metrology promises high-precision measurements of classical
parameters with far reaching implications for science and
technology. 
So far, research has concentrated almost exclusively on 
quantum-enhancements in integrable systems, such as 
precessing spins  or harmonic 
oscillators prepared in non-classical states. Here we show that large
benefits can be drawn from 
rendering integrable quantum sensors chaotic, both in terms of 
achievable sensitivity as well as robustness to noise, while avoiding
the challenge of preparing and protecting large-scale entanglement. We
apply the method to spin-precession magnetometry
and show in particular that the sensitivity of state-of-the-art
magnetometers can be further enhanced by
subjecting the spin-precession to non-linear kicks that renders the
dynamics chaotic.
\end{abstract}

\maketitle

\section*{Introduction}
Quantum-enhanced measurements (QEM) use quantum effects in order to
measure physical quantities with larger precision than what is
possible classically with comparable resources. QEMs 
are therefore expected to have large impact in many areas, such as
improvement  of frequency standards
\cite{Huelga97,PhysRevLett.86.5870,Leibfried04,wasilewski2010quantum,koschorreck2010sub},   
gravitational wave detection 
\cite{Goda08,aasi2013enhanced},                               
navigation 
\cite{Giovannetti01}, remote sensing \cite{Allen08radar}, or measurement
of very small magnetic 
fields \cite{Taylor08}.    
A well known example is the use of so-called
NOON states in an interferometer, where a state with $N$ photons in one
arm of the interferometer and zero in the other is superposed with the
opposite situation \cite{Boto00}.  It was shown that the smallest phase shift
that such an interferometer could measure scales as $1/N$, a large
improvement over the standard $1/\sqrt{N}$ behavior that one obtains from
ordinary laser light.  The latter scaling is known as the standard
quantum limit (SQL), and the $1/N$ scaling as the
Heisenberg limit (HL). So far the SQL has been beaten only in few
experiments, and only for small $N$ (see 
e.g.~\cite{Higgins07,Leibfried04,Nagata07}),
as the required non-classical states are
difficult to prepare and stabilize and are prone to
decoherence.\\
Sensing devices used in quantum metrology so far have been based
almost exclusively on integrable systems, such as precessing spins
(e.g.~nuclear spins, NVcenters, 
etc.) or harmonic oscillators (e.g.~modes of an electro-magnetic field
or mechanical oscillators),  
prepared in non-classical states (see \cite{pezze_non-classical_2016}
for a recent 
review). The idea of 
the present work is to 
achieve enhanced measurement precision with readily accessible input
states by disrupting the parameter coding by a sequence of controlled
pulses that render the dynamics chaotic. At first sight this may 
appear a bad idea, as measuring something precisely requires
well-defined, reproducible behavior, whereas classical chaos is
associated with unpredictible long-term behavior. However, the extreme
sensitivity to initial conditions underlying classically chaotic
behavior is absent in the quantum world with its unitary dynamics 
in Hilbert space that preserves distances between 
states. In turn, quantum-chaotic dynamics can lead to
exponential sensitivity with respect to parameters of the system
\cite{Peres91}.  \\
The sensitivity to changes of a parameter of quantum-chaotic systems has
been studied in great detail with the technique 
of Loschmidt echo \cite{Gorin06review}, which measures the overlap
between a state propagated forward with a unitary operator
and propagated 
backward with a slightly perturbed unitary operator. In the limit of
infinitesimally small perturbation, the Loschmidt
echo turns out to be directly related to the quantum Fisher
information (QFI) that determines the smallest uncertainty with
which a parameter can be estimated. 
Hence, a wealth of known results from quantum chaos can be immediately
translated to study the ultimate sensitivity of quantum-chaotic
sensors. 
In particular, linear response expressions for fidelity
can be directly transfered to the {exact} expressions for
the QFI.\\
Ideas of replacing entanglement creation by dynamics were proposed
previously
\cite{Boixo08.2,xiao2011chaos,song2012quantum,weiss2009signatures,frowis2016detecting},
but focussed on initial state preparation, or robustness of the readout
\cite{macri2016loschmidt,linnemann2016quantum}, without 
introducing or exploiting chaotic dynamics during the
parameter encoding.  They are hence comparable to  
spin-squeezing of the input 
state \cite{ma2011quantum}.
Quantum chaos is also favorable for state tomography of random initial
states with weak continuous 
time measurement  \cite{madhok2014information,madhok2016review}, but
no attempt was made to use this for precision 
measurements of a parameter.
A recent review of other approaches to quantum-enhanced metrology
that avoid initial entanglement can be found in 
\cite{braun2017without}.\\
We study  quantum-chaotic enhancement of sensitivity at the example
of the measurement of a classical magnetic field with a
spin-precession magnetometer.  In these
devices that count amongst the most 
sensitive magnetometers currently available
\cite{allred2002high,Kominis03,savukov2005effects,budker2007optical,sheng2013subfemtotesla},
the magnetic field is coded in a precession frequency of  atomic
spins that act 
as the sensor.  We show that the 
precision of the magnetic-field measurement can be substantially
enhanced by non-linearly kicking the spin 
during the precession phase and driving it into a 
chaotic regime. The initial state can be chosen as an
essentially classical state, in particular a state without initial
entanglement.  The enhancement is robust with
respect to decoherence or dissipation. We demonstrate this by modeling
the magnetometer on two different levels: 
firstly as a kicked top, a well-known system in quantum chaos to
which we add dissipation through superradiant damping; and
secondly with a detailed realistic model of a
spin-exchange-relaxation-free atom-vapor magnetometer including all
relevant decoherence mechanisms
\cite{allred2002high,appelt1998theory},
to which we add non-linear kicks.  
\section*{Results}
\subsection*{Physical model of a quantum-chaotic sensor}
As a sensor we consider a kicked top
(KT), a well-studied quantum-chaotic system 
\cite{haake1987classical,kickedtop,haake2013quantum} described by the
time-dependent Hamiltonian  
\begin{equation}\label{eq:hamilton}
{H}_\text{KT}(t)=\alpha J_z+\frac{k}{2J}J_y^2\sum_{n=-\infty}^\infty \tau\delta(t-n\tau)\,,
\end{equation}
where $J_i$ ($i=x,y,z$) are components of the (pseudo-)angular momentum
operator, 
$J\equiv j+1/2$, and we set $\hbar=1$. 
$J_z$ generates a precession of the (pseudo-)angular momentum
vector about the $z$-axis with precession
angle $\alpha$ which is the parameter we want to estimate. ``Pseudo''
refers to the fact that the physical system need not be an actual
physical spin, but can be any system with $2j+1$ basis states on which
the $J_i$ act accordingly.  For a
physical spin-$j$ in a magnetic field $B$ in $z$-direction, $\alpha$
is directly 
proportional to $B$. The $J_y^2$-term is the non-linearity, 
assumed to act instantaneously compared to the precession, controlled
by the kicking strength $k$ and  applied periodically with a period
$\tau$ that leads to chaotic behavior. The 
system can be described stroboscopically with discrete time $t$ in
units of $\tau$ (set to $\tau=1$ in the following),
\begin{equation}  \label{eq:Upsi}
\ket{\psi(t)}=U_\alpha(k)\ket{\psi(t-1)}=U^t_\alpha(k)\ket{\psi(0)}
\end{equation}
with  the unitary Floquet-operator
\begin{equation}
U_\alpha(k)=T\exp\left(-i\int_{t}^{t+1}\ dt' 
{H}_\text{KT}(t')\right)=e^{-ik\frac{ J^2_y
	}{2J}}e^{-i\alpha J_z}
\end{equation}
that propagates the state of the 
system from right after a kick to right after the 
next kick \cite{haake1987classical,kickedtop,haake2013quantum}. $T$
denotes time-ordering.
The total spin is conserved, and $1/J$ can be identified with an 
effective $\hbar$, such that the 
limit $j\to\infty$ corresponds to the classical limit, where $X=J_x/J$,
$Y=J_y/J$, $Z=J_z/J$ become 
classical variables confined to the unit sphere.
$(Z,\phi)$ can be identified with classical phase
space variables, where $\phi$ is the azimuthal angle of
$\textbf{X}=(X,Y,Z)$ \cite{haake2013quantum}. 
For $k=0$ the dynamics is integrable,
as the precession 
conserves $Z$ and increases $\phi$ by $\alpha$ 
for each application of
$U_\alpha(0)$. Phase space portraits of the corresponding classical
map show that for $k\lesssim 2.5$, the dynamics remains close to
integrable with 
large visible Kolmogorov-Arnold-Moser tori, whereas for $k\gtrsim 3.0$ the chaotic dynamics 
dominates \cite{haake2013quantum}.

\begin{figure}[h]
		\includegraphics[width=9cm]{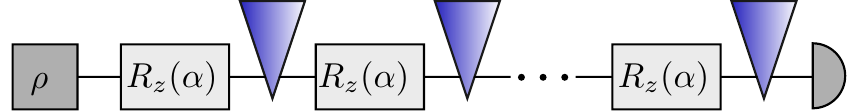}
	\caption{Schematic representation of the parameter encoding: 
		Propagation starts on the left with an initial state $\rho$ and ends on the right with a measurement (semi-circle symbol). The encoding through linear precession $R_z(\alpha)$ 
		about the $z$-axis by an angle $\alpha$ 
		is periodically disrupted through
		parameter independent, non-linear, controlled kicks (blue triangles) that can
		render the system chaotic.}\label{fig:scheme}
\end{figure}

States that correspond most closely to classical phase space points
located at $(\theta,\phi)$ are SU(2)-coherent states
(``spin-coherent states'', or ``coherent states'' for short), defined as  
\begin{equation}\label{eq:coherent}
\ket{j,\theta,\phi}=\sum_{m=-j}^j\sqrt{\binom{2j}{j-m}}\sin(\theta/2)^{j-m}\cos(\theta/2)^{j+m}e^{i(j-m)\phi}\ket{jm}
\end{equation}
in the usual notation of angular momentum states $\ket{jm}$
(eigenbasis of ${\bf J}^2$ and $J_z$ with eigenvalues $j(j+1)$ and 
$m$, $2j\in \mathbb N$, $m=-j,-j+1,\ldots,j$). They are localized at
polar and azimuthal 
angles $\theta,\phi$ with smallest possible uncertainty of all
spin-$j$ states (associated circular area $\sim 1/j$ in phase space). They remain
coherent states under the action of $U_\alpha(0)$, i.e.~just
get rotated, $\phi\mapsto\phi+\alpha$.
For the KT, the parameter encoding of $\alpha$ in the
quantum state breaks with the standard encoding scheme (initial state
preparation, parameter-dependent precession, measurement)  
by periodically disrupting the coding evolution
with parameter-independent kicks that generate chaotic behavior (see
Fig.~\ref{fig:scheme}).

An experimental realization of the
kicked top was proposed in \cite{haake2000can}, including superradiant
dissipation.  It has been realized
experimentally \cite{chaudhury2009quantum} 
in cold cesium 
 vapor using optical pulses
 (see Supplementary Note 1 for details).

\subsection*{Quantum parameter estimation theory}
Quantum  measurements are most conveniently described by a positive-operator
valued measure (POVM) $\{\Pi_\xi\}$ with positive operators
$\Pi_\xi$ (POVM-elements) that fulfill $\int d\xi \Pi_\xi=\bOne$. 
Measuring a quantum state described by a density 
operator $\rho_\alpha$ yields for a given POVM and a given parameter
$\alpha$ encoded in the quantum state 
 a probability distribution
$p_\alpha(\xi)=\tr(\Pi_\xi\rho_\alpha)$ of measurement results 
$\xi$. The Fisher information $I_{\text{Fisher},\alpha}$ is then 
defined by
\begin{equation}\label{eq:Fisher}
I_{\text{Fisher},\alpha}:=\int d\xi\frac{(dp_\alpha(\xi)/d\alpha)^2}{p_\alpha(\xi)}\,. 
\end{equation}
The  minimal achievable uncertainty, i.e.~the variance 
of the estimator $ \text{Var}(\alpha_\text{est})$, with
which a parameter $\alpha$ of a state $\rho_\alpha$ can 
be estimated {for a given POVM} 
with $M$ independent 
measurements is given by the Cram\'er-Rao bound,
$ \text{Var}(\alpha_\text{est}) \geq 1/(M
I_{\text{Fisher},\alpha})$. 
Further optimization over all possible (POVM-)measurements 
leads to the quantum-Cram\'er-Rao bound (QCRB),  
\begin{equation}\label{eq:qfi_inegality}
\text{Var}(\alpha_\text{est}) \geq \frac{1}{M I_\alpha},
\end{equation} 
which presents an ultimate bound on the minimal 
achievable uncertainty, where $I_\alpha$ is the quantum Fisher
information (QFI), and $M$ the 
number of independent measurements \cite{Helstrom1969}.

The QFI is related to the Bures distance $ds_{\rm Bures}^2$
between
the states $\rho_\alpha$ and $\rho_{\alpha+d\alpha}$, separated by an
infinitesimal change of the parameter $\alpha$, 
$ds_{\rm Bures}^2(\rho,\sigma)\equiv
2\left(1-\sqrt{F(\rho,\sigma)}\right)$.
The fidelity $F(\rho,\sigma)$ is defined as
$F(\rho,\sigma)=||\rho^{1/2}\sigma^{1/2}||^2_1$, and $||A||_1\equiv {\rm
tr}\sqrt{AA^\dagger}$ denotes the trace norm \cite{Miszczak09}. With
this,
\begin{equation}
  \label{eq:QFI}
I_\alpha=4ds_{\rm
Bures}^2(\rho_\alpha,\rho_{\alpha+d\alpha})/d\alpha^2  
\end{equation}
\cite{Braunstein94}.
For pure states
$\rho=\ket{\psi}\bra{\psi}$, $\sigma=\ket{\phi}\bra{\phi}$, the fidelity is
simply given by $F(\rho,\sigma)=|\braket{\psi|\phi}|^2$. A parameter
coded in a pure state via the unitary transformation
$\ket{\psi_\alpha}=e^{-i\alpha G}\ket{\psi(0)}$ with hermitian
generator $G$ gives the QFI \cite{Braunstein90}
\begin{equation}\label{eq:IG}
I_\alpha=4\,\text{Var}(G)\equiv  4(\braket{G^2}-\braket{G}^2)\,,
\end{equation} 
which holds for all $\alpha$, and
where $\braket{\cdot}\equiv\bra{\psi_\alpha}\cdot\ket{\psi_\alpha}$.

\subsection*{Loschmidt echo}
The sensitivity to changes of a parameter of quantum-chaotic systems has
been studied in great detail with the technique 
of Loschmidt echo \cite{Gorin06review}, which measures the overlap
$F_\epsilon(t)$ 
between a state propagated forward with a unitary 
operator $U_\alpha(t)$ and propagated
backward with a slightly perturbed unitary operator
$U_{\alpha+\epsilon}(-t)=U_{\alpha+\epsilon}^\dagger(t)$, where
$U_\alpha(t)=T\exp\left(-\frac{i}{\hbar}\int_0^t
dt'H_\alpha(t')\right)$ {with the time ordering operator $T$, the
Hamiltonian}, $H_{\alpha+\epsilon}(t)=H_\alpha(t)+\epsilon V(t)$ 
and the perturbation $V(t)$,
\begin{equation}\label{eq:fidelity}
F_\epsilon(t)=|\braket{\psi(0)|U_{\alpha}(t)U_{\alpha+\epsilon}(-t)\psi(0)}|^2\, \,.
\end{equation}
$F_\epsilon$ is exactly the fidelity that enters via the Bures distance in the definition
eq.~(\ref{eq:QFI}) of 
the QFI for pure states, such that 
$I_\alpha (t)=\lim_{\epsilon\to 0}4\frac{1-F_\epsilon(t)}{\epsilon^2}$. 

\subsection*{Benchmarks}
In order to assess the
influence of the kicking on the QFI, we calculate as benchmarks the
QFI for the (integrable) top with Floquet operator $U_\alpha(0)$
without kicking, both for an initial coherent state and for a Greenberger-Horne-Zeilinger
(GHZ) state $\ket{\psi_\text{GHZ}}=(\ket{j,j}+\ket{j,-j})/\sqrt{2}$. The latter 
is
the equivalent of a NOON state written in terms of (pseudo-)angular
momentum states. 
The QFI for the time evolution eq.~\eqref{eq:Upsi} of a top with  
Floquet operator $U_\alpha(0)$
is  given 
by eq.~\eqref{eq:IG} with $G=J_z$. For an initial
coherent state located at $\theta,\phi$   it results in a QFI
\begin{equation}
\label{eq:top}
I_\alpha(t)=2t^2j\sin^2\theta.
\end{equation}
As expected, $I_\alpha(t)=0$ for $\theta=0$ where the coherent
state is an eigenstate of $U_\alpha(0)$.  The scaling
$\propto t^2$ is typical of quantum coherence, and $I_\alpha(t)\propto
j$ signifies a SQL-type scaling with $N=2j$, when 
the spin-$j$ is composed of $N$ spin-$\frac{1}{2}$ particles in
a state invariant under permutations of particles. 
 For the benchmark we use the
optimal value $\theta=\pi/2$ in eq.~\eqref{eq:top}, 
i.e.~$ I_\text{top,CS}\equiv 2t^2j$. 
For a GHZ state, the QFI becomes 
\begin{equation}
\label{eq:ghz}
I_\alpha(t)=4t^2j^2\equiv I_\text{top,GHZ}\,,
\end{equation}
which clearly displays the HL-type scaling 
$\propto(2j)^2\equiv N^2$.

\subsection*{Results for the kicked top without dissipation}
In the fully 
chaotic case, known results for the Loschmidt echo
suggest a QFI of  the KT $\propto t j^2$ for times $t$ with
$t_E<t<t_H$, where 
$t_\text{E}=\frac{1}{\lambda}\ln\left(\frac{\Omega_\text{V}}{h^{d}}\right)$ is the
Ehrenfest time, and $t_\text{H}=\hbar/\Delta$  the    
 Heisenberg time; 
$\lambda$ is the Lyapunov exponent,  $\Omega_\text{V}$ the volume of
phase-space, $h^d$ with $d$ the number of degrees of freedom the
volume of a Planck cell, and $\Delta$ the mean
 energy level spacing
 \cite{zaslavsky1981stochasticity,haake2013quantum,Gorin06review}.
 For the kicked top, $h^d\simeq \Omega_\text{V}/(2J)$.
 More precisely, we find for $t\simeq t_E$ a QFI $I_\alpha\propto tj^2$ 
and for $t\gg t_H$ (see Methods)
\begin{equation}\label{eq:fidelity:chaos:Heisenberg}
   I(t)=8s\sigma_\text{cl}t^2J\,,
\end{equation}
where $s$ denotes the number of invariant
subspaces $s$ of the Hilbert 
space ($s=3$ for the kicked top with $\alpha=\pi/2$, see page 359 in
\cite{Peres91}), and 
$\sigma_\text{cl}$ is a transport coefficient that can be calculated 
numerically.  The infinitesimally small
perturbation relevant for the QFI  makes that one is always in
the perturbative regime 
\cite{PhysRevE.64.055203,PhysRevE.65.066205}. The Gaussian decay of Loschmidt echo
characteristic of that regime becomes 
the slower the smaller the perturbation and goes over
into a power law in the limit of infinitesimally small perturbation
\cite{Gorin06review}.

\begin{figure}
		\includegraphics[width=5.6666cm]{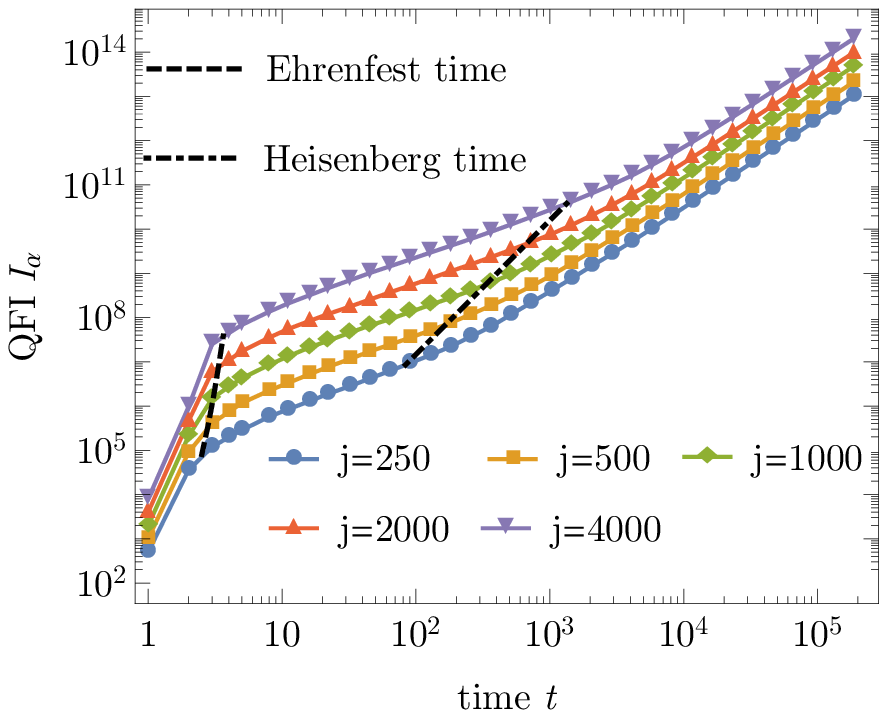}
		\hspace{0.5cm}
		\includegraphics[width=5.6666cm]{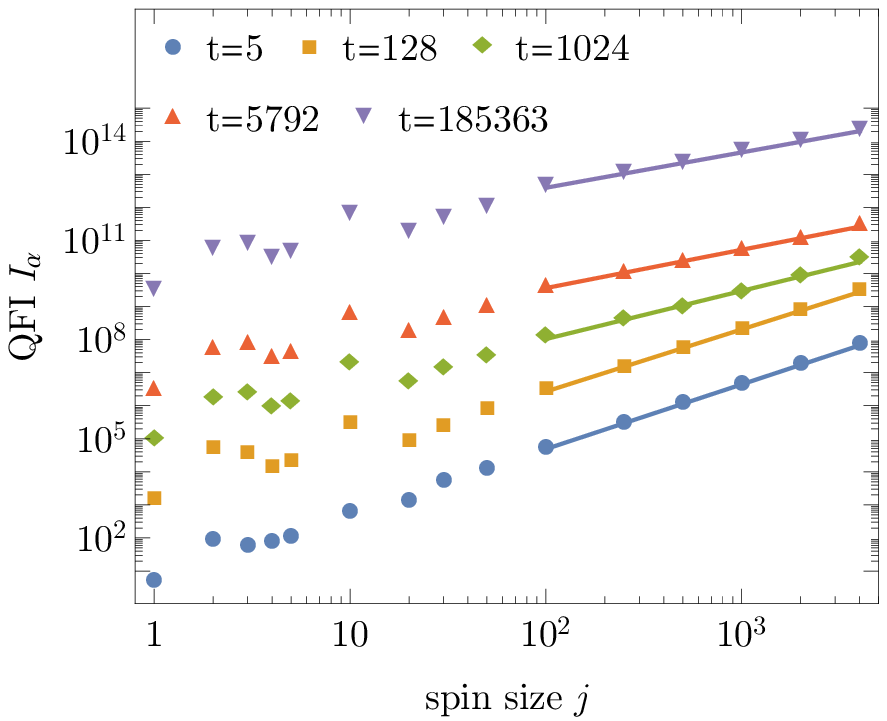}
		\hspace{0.5cm}
		\includegraphics[width=5.6666cm]{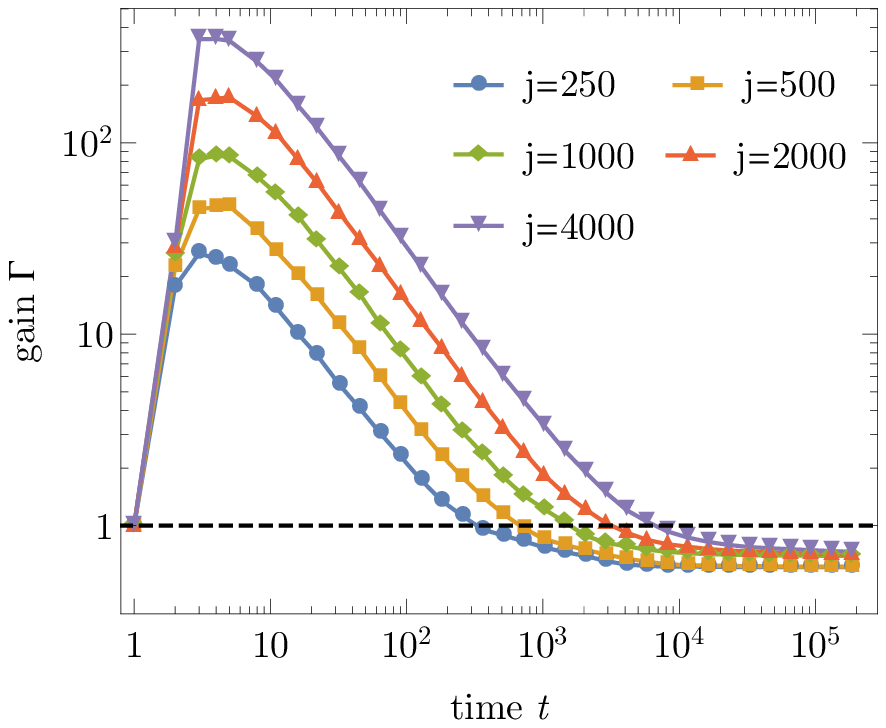}\\
		\hspace{0.5cm} {\bf (a)}\hspace{5.7cm}{\bf (b)}\hspace{5.8cm}{\bf (c)}
	\caption{Quantum-chaotic 
		enhancement of sensitivity. 
		{\bf (a)} 
		$t$-scaling of the quantum Fisher information (QFI) $I_\alpha$ in the strongly chaotic case. Dashed and 
		dash-dotted lines indicate Ehrenfest and Heisenberg times, respectively. 
		{\bf (b)}
		$j$-scaling of the QFI. Fits have slopes
		$1.96,1.88,1.46,1.16$, and $1.08$ in increasing order of $t$. 
		{\bf (c)}
		Gain $\Gamma=I_{\alpha,\textrm{KT}}/I_\text{top,CS}$ 
		as function of time $t$ for different values of spin size $j$. 
		The dashed black line marks the threshold
		$\Gamma=1$. Kicking strength $k=30$, and initial coherent state at
		$(\theta,\phi)=(\pi/2,\pi/2)$ in all plots. }\label{fig:QFI_scal_k30} 
\end{figure}
The numerical results for the QFI in Fig.~\ref{fig:QFI_scal_k30} illustrate a
cross-over of power-law scalings in the fully
chaotic case ($k=30$) for an initial coherent state located on the
equator $(\theta,\phi)=(\pi/2,\pi/2)$.  
The analytical Loschmidt echo results are nicely 
reproduced: A smooth transition in scaling 
from $tj^2\rightarrow t^2j$ for 
$t=t_\text{E}\rightarrow t\gtrsim t_\text{H}$ can be observed and 
confirms
eqs.~(\ref{eq:qfi:chaos:ehrenfest}),(\ref{eq:qfi:chaos:ehrenfest:1})
in the Methods
for $t>t_\text{E}=\ln(2J)/\lambda$, with the numerically determined 
Lyapunov exponent 
$\lambda\simeq 2.4733$, and eq.~(\ref{eq:fidelity:chaos:Heisenberg}) 
for $t\gtrsim t_\text{H}\simeq J/3$ \cite{Gorin06review}. We find for
relatively large $j$ ($j\gtrsim 10^2$) a scaling 
$I_\alpha\propto j^{1.08}$ in good agreement with
eq.~(\ref{eq:fidelity:chaos:Heisenberg}) predicting a linear
$j$-dependence  for large $t\gtrsim t_\text{H}$. During the transient time
$t<t_\text{E}$, when the state  
is spread over the phase space, QFI shows a rapid growth that can be
attributed to the generation of  
coherences that are particularly sensitive to the precession.

The comparison of the KT's QFI $I_{\alpha,\textrm{KT}}$ with 
the benchmark $I_\text{top,CS}$ of the 
integrable top in Fig.~\ref{fig:QFI_scal_k30} \textbf{(c)} shows
that a gain of more than two orders
of magnitude  for 
$j=4000$ can be found at $t\lesssim  t_\text{E}$. Around $t_\text{E}$
the state has spread over the phase  
space  and has developed coherences while for larger times
$t>t_\text{E}$ the top  
catches up due to its superior time scaling ($t^2$ vs.~$t$). The
long-time behavior yields a constant gain less than 1, which means
that the top achieves a higher QFI than the KT in this
regime. The gain 
becomes constant as both top and 
KT  exhibit a $t^2$ scaling of the QFI.

\begin{figure}
	\includegraphics[width=5.6666cm]{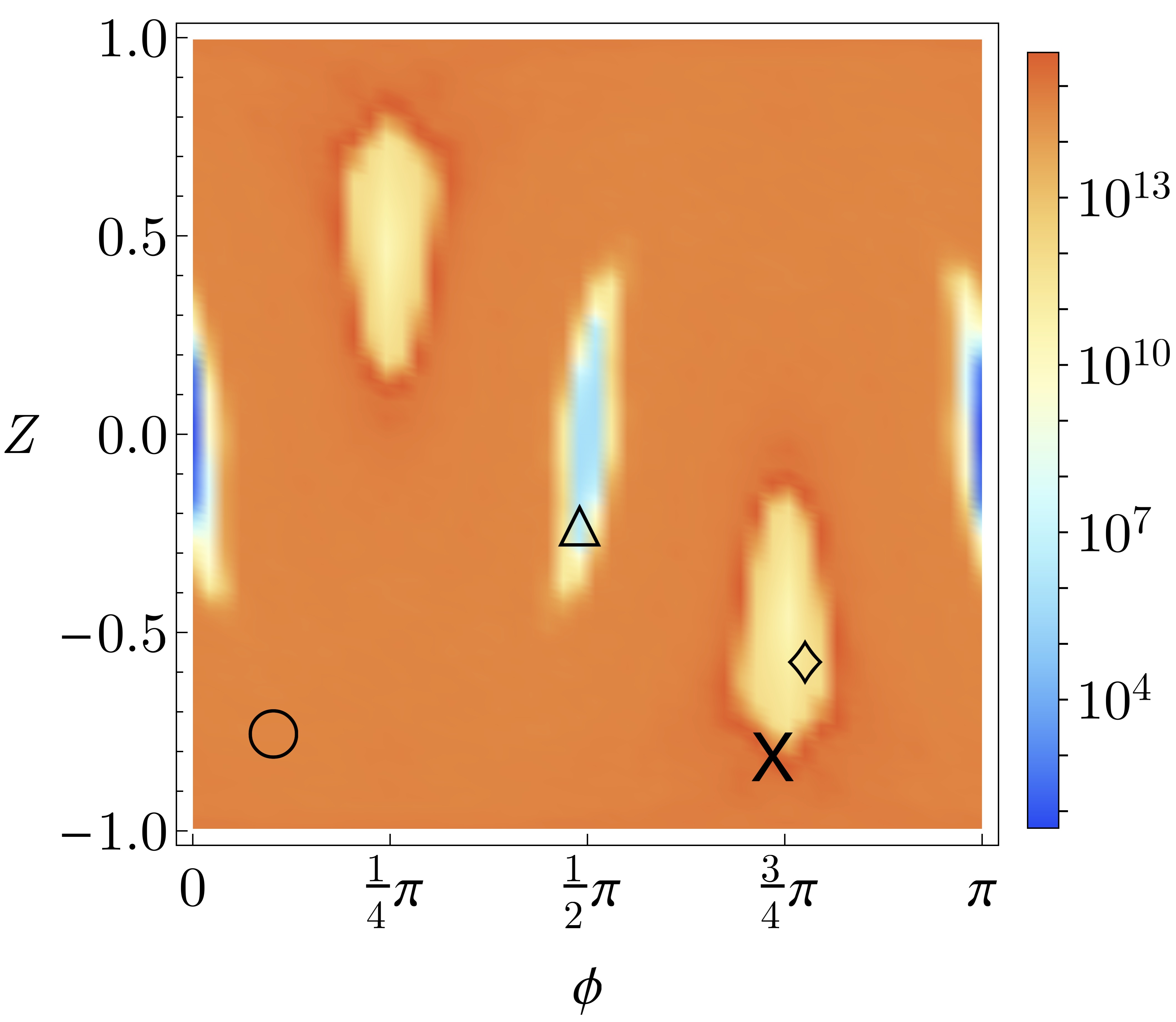}
	\hspace{0.5cm}
	\includegraphics[width=5.6666cm]{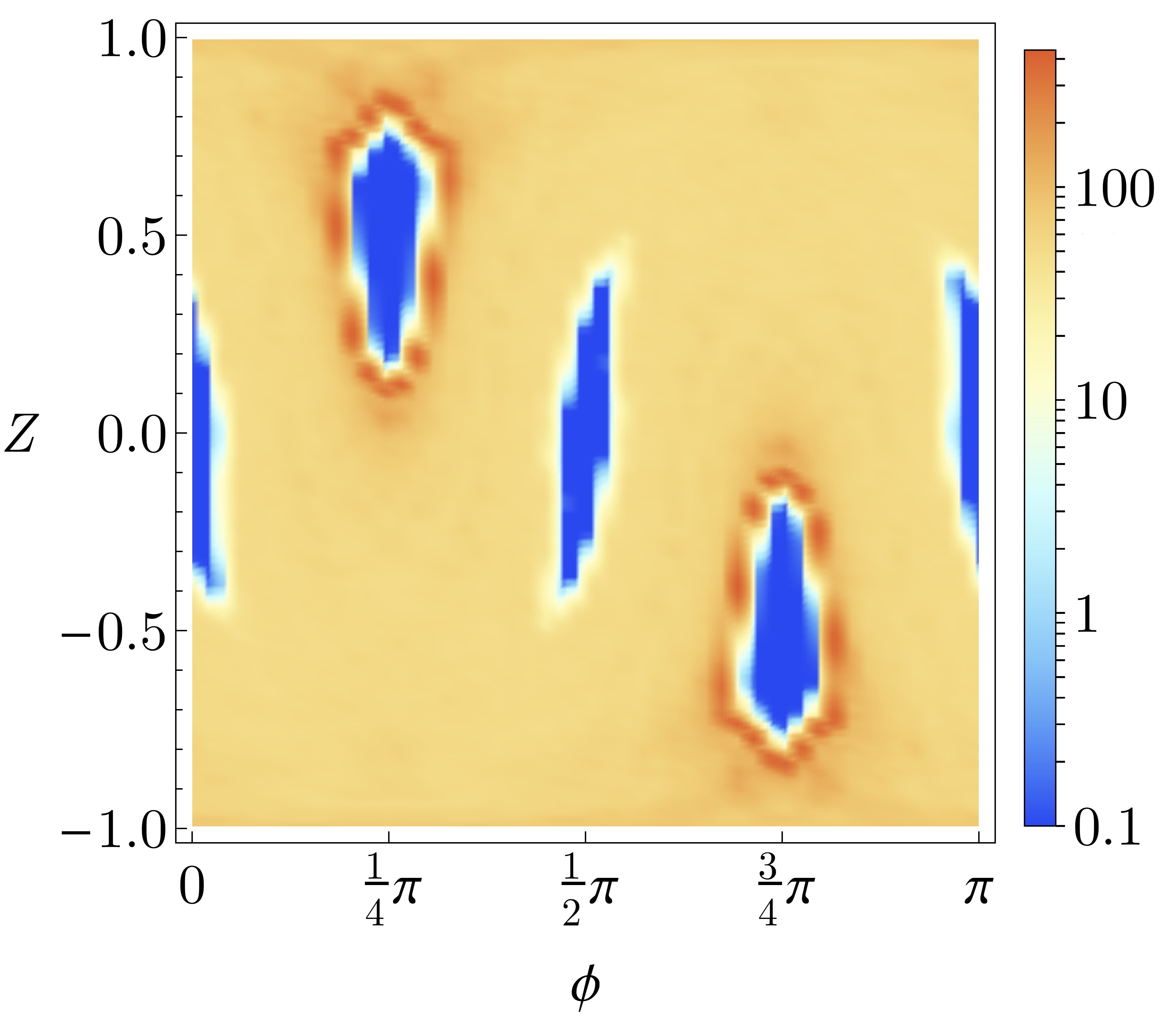}
	\hspace{0.5cm}
	\includegraphics[width=5.6cm]{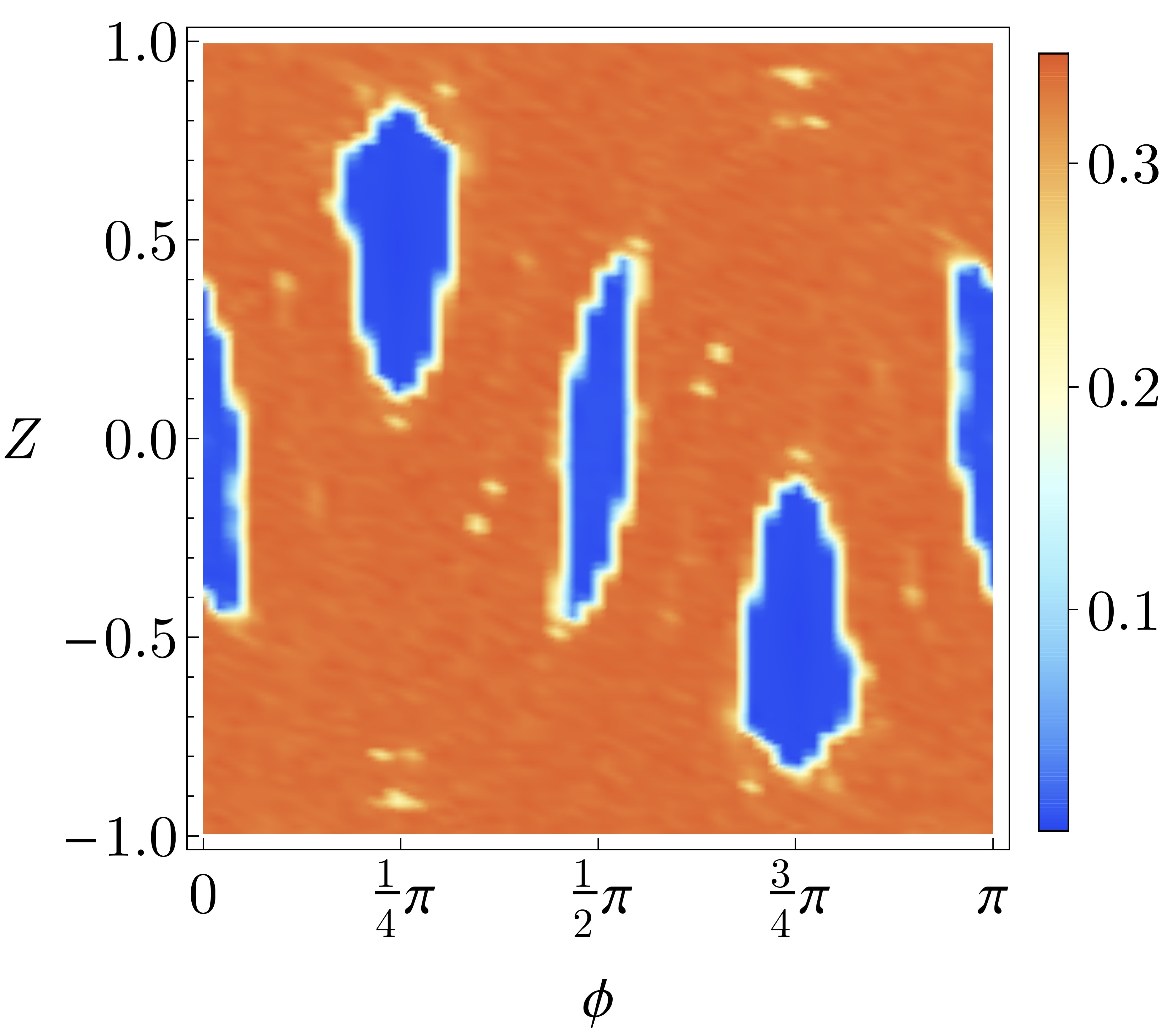}
	
	{\bf (a)}\hspace{5.9cm}{\bf (b)}\hspace{5.9cm}{\bf (c)}
	\caption{Portraits of a mixed phase space for large spin size $j=4000$. 
		The	heat maps depict quantum Fisher information (QFI) in panel {\bf (a)} and gain $\Gamma$ in panel {\bf (b)} 
		at time $t=2^{15}$ and kicking strength $k=3$. $\Gamma<0.1$ is 
		depicted as $\Gamma=0.1$ to simplify the representation. 
		A gain of more than two orders of
		magnitude in the QFI is observed for edge states, localized on the
		border of the stability islands. For comparison, panel {\bf (c)} 
		depicts the classical Lyapunov exponent $\lambda$. }\label{fig:QFI_heat} 
\end{figure}
Whereas in the fully chaotic regime the memory of the initial state is
rapidly forgotten, and the initial state can therefore be chosen
anywhere in the chaotic sea without changing much the QFI,  the
situation is very different in the case of a mixed phase space, in
which stability islands are still present.  
Fig.~\ref{fig:QFI_heat} shows phase space distributions for $k=3$ of
QFI and gain $\Gamma=I_{\alpha,\textrm{KT}}/I_\text{top,CS}$ 
exemplarily for large $t$ and $j$ ($t=2^{15}$, $j=4000$) in
comparison with the classical Lyapunov exponent, where $\phi,Z$
signify the position of the initial coherent state. 

The QFI nicely 
reproduces the essential 
structure in classical phase space for the Lyapunov exponent $\lambda$
(see Methods for the calculation of $\lambda$). 
Outside the 
regions of classically regular motion, the KT
clearly outperforms  
the top by more than 
two orders of magnitude. Remarkably, the QFI is highest 
at the  
boundary of non-equatorial islands of classically regular
motion. Coherent states located on that boundary will be called
edge states. \\

\begin{figure}
		\includegraphics[width=8.75cm]{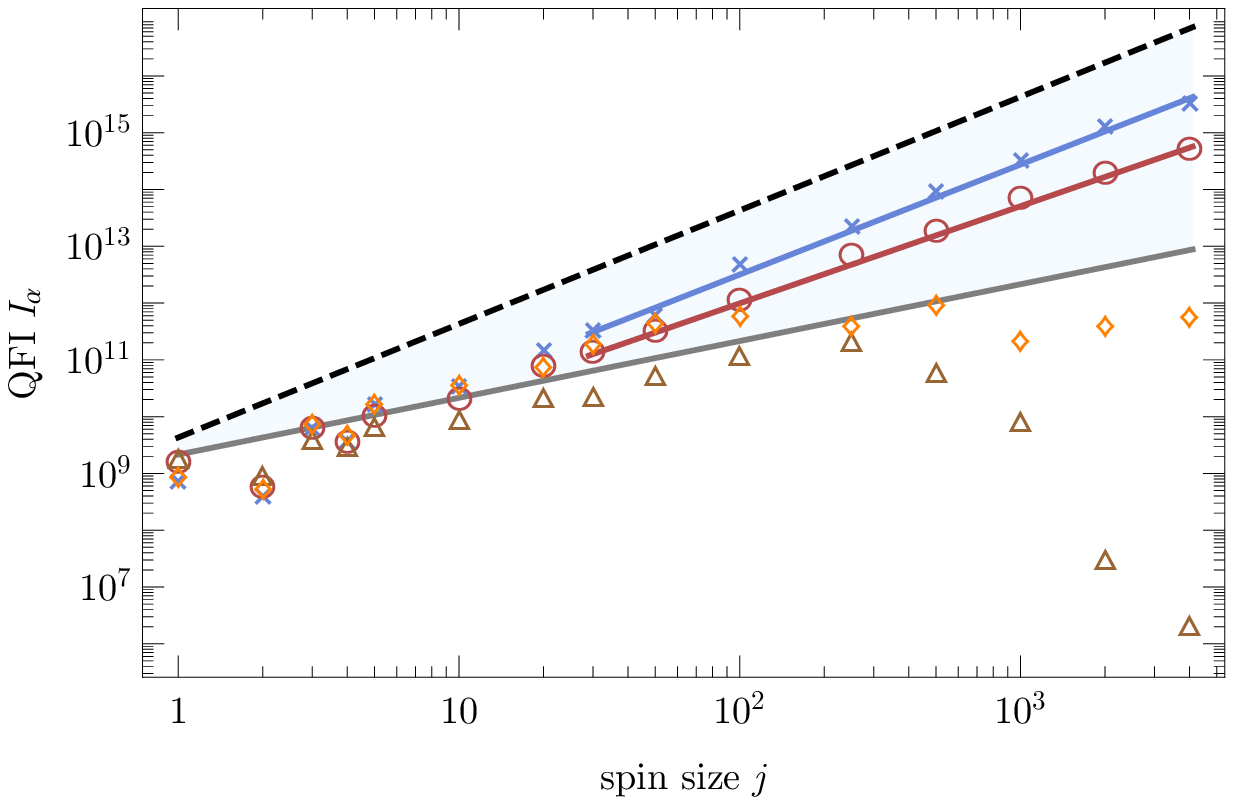}
		\hspace{0.5cm}
		\includegraphics[width=8.75cm]{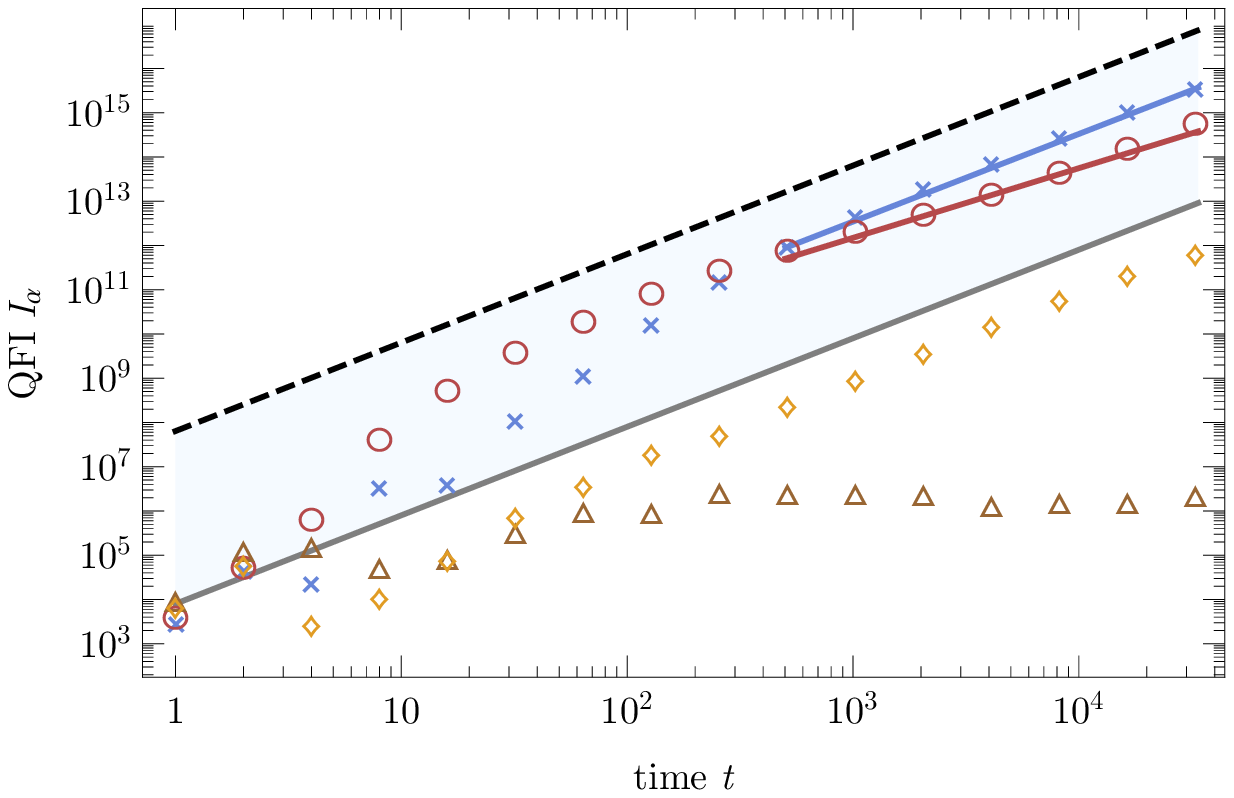}\\
		\hspace{0.8cm}\textbf{(a)}\hspace{9cm}\textbf{(b)}
	\caption{Scaling of quantum Fisher information (QFI) in different areas of a mixed phase space. 
		QFI $I_\alpha(j)$ 
		(panel {\bf (a)} for time $t=2^{15}$) and $I_\alpha(t)$ (panel {\bf (b)} for
		spin size $j=4000$) for kicking strength $k=3$ and different initial coherent states $\ket{\theta,\phi}$
		(as marked in panel {\bf (a)} of Figure~\ref{fig:QFI_heat}): inside an
		equatorial (brown triangles,  $\ket{1.82,1.54}$) and a
		non-equatorial (orange diamonds, $\ket{2.20,2.44}$) stability
		island, in the chaotic sea (red circles, $\ket{2.46,0.32}$), and an
		edge state (blue crosses, $\ket{2.56,2.31}$). Benchmarks 
		$I_\text{top,CS}$ (gray line) and 
		$I_\text{top,GHZ}$ (black dashed line) represent the standard quantum limit 
		and Heisenberg limit, respectively. Fits for the edge state and the 
		state in the chaotic sea exhibit a slope of $1.94$ and $1.71$ for
		the $j$-scaling, and $1.98$ and $1.57$ for the
		$t$-scaling.}\label{fig:KTmix_scaling_compact} 
\end{figure}
The diverse dynamics for different 
phase space 
regions  calls for   dedicated  
analyses. 
Fig.~\ref{fig:KTmix_scaling_compact}
depicts the QFI with respect to $j$ 
and $t$. 
The blue area is
lower and upper bounded by the benchmarks 
$I_\text{top,CS}$ and $I_\text{top,GHZ}$,
eq.~(\ref{eq:ghz}), respectively.  

\begin{figure}
		\includegraphics[width=8.75cm]{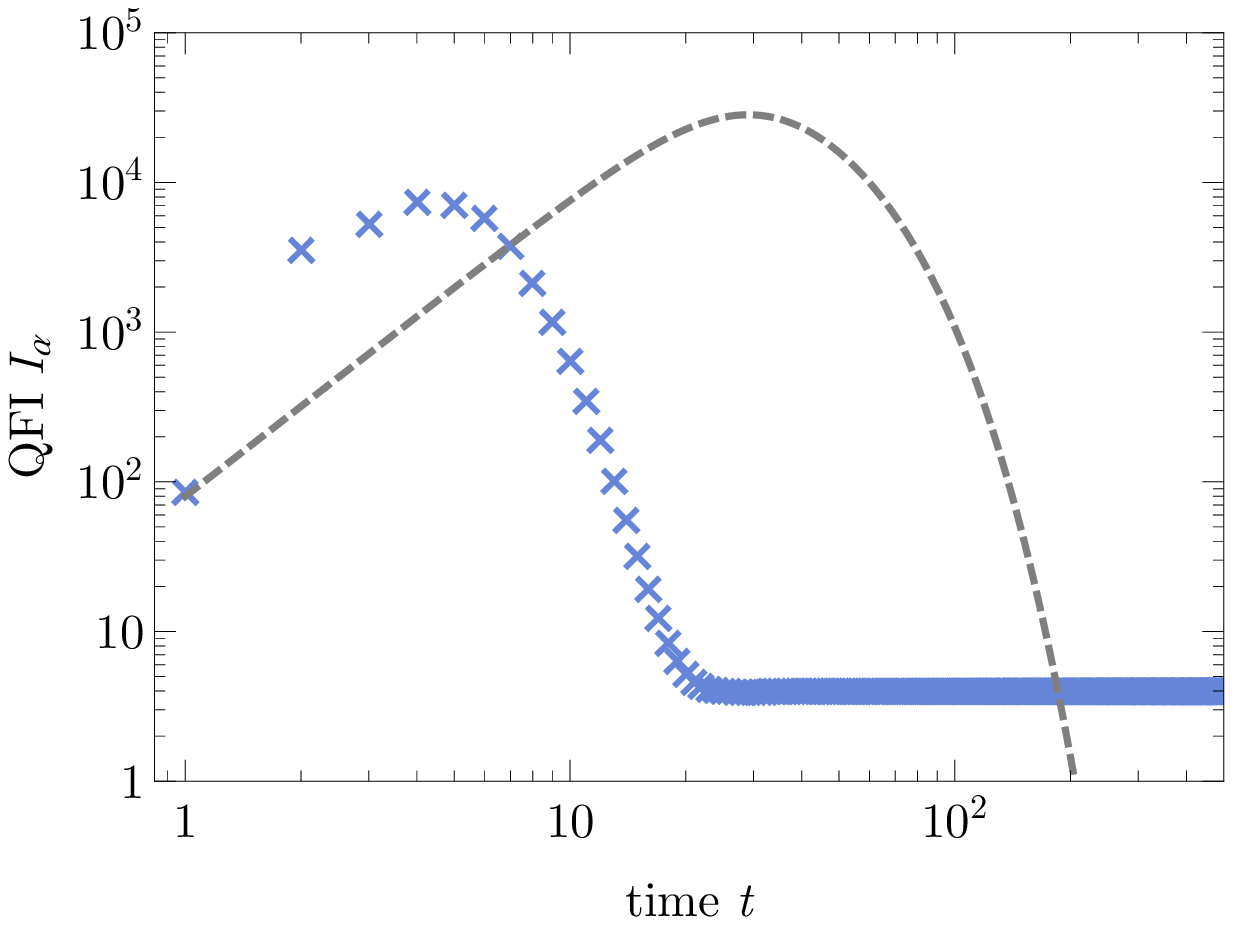}\hspace{0.5cm}
		\includegraphics[width=8.75cm]{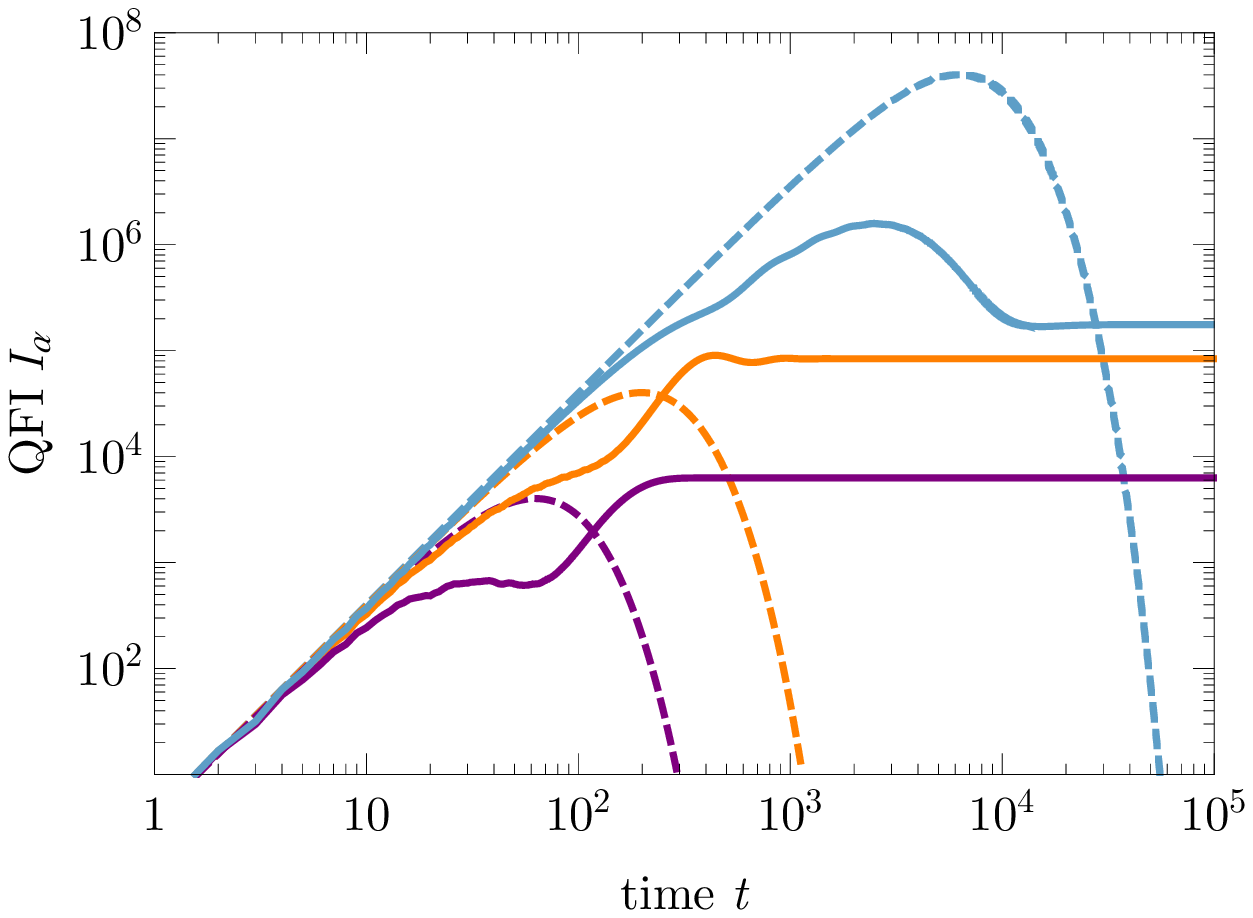}
		\\
		{\bf (a)}\hspace{8.75cm}{ \bf (b)}\\
		\vspace{1cm}
		\includegraphics[width=8.75cm]{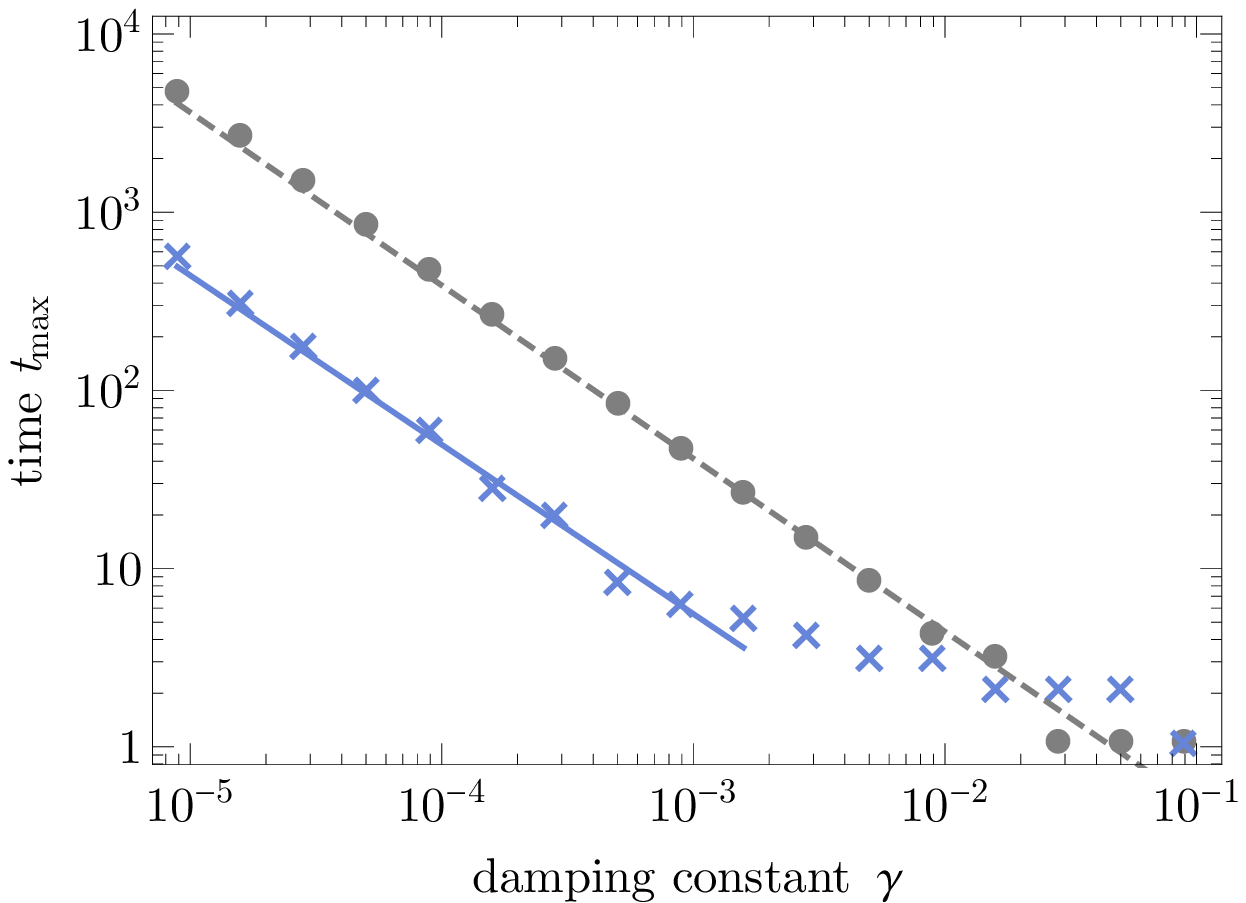}\hspace{0.5cm}
		\includegraphics[width=8.75cm]{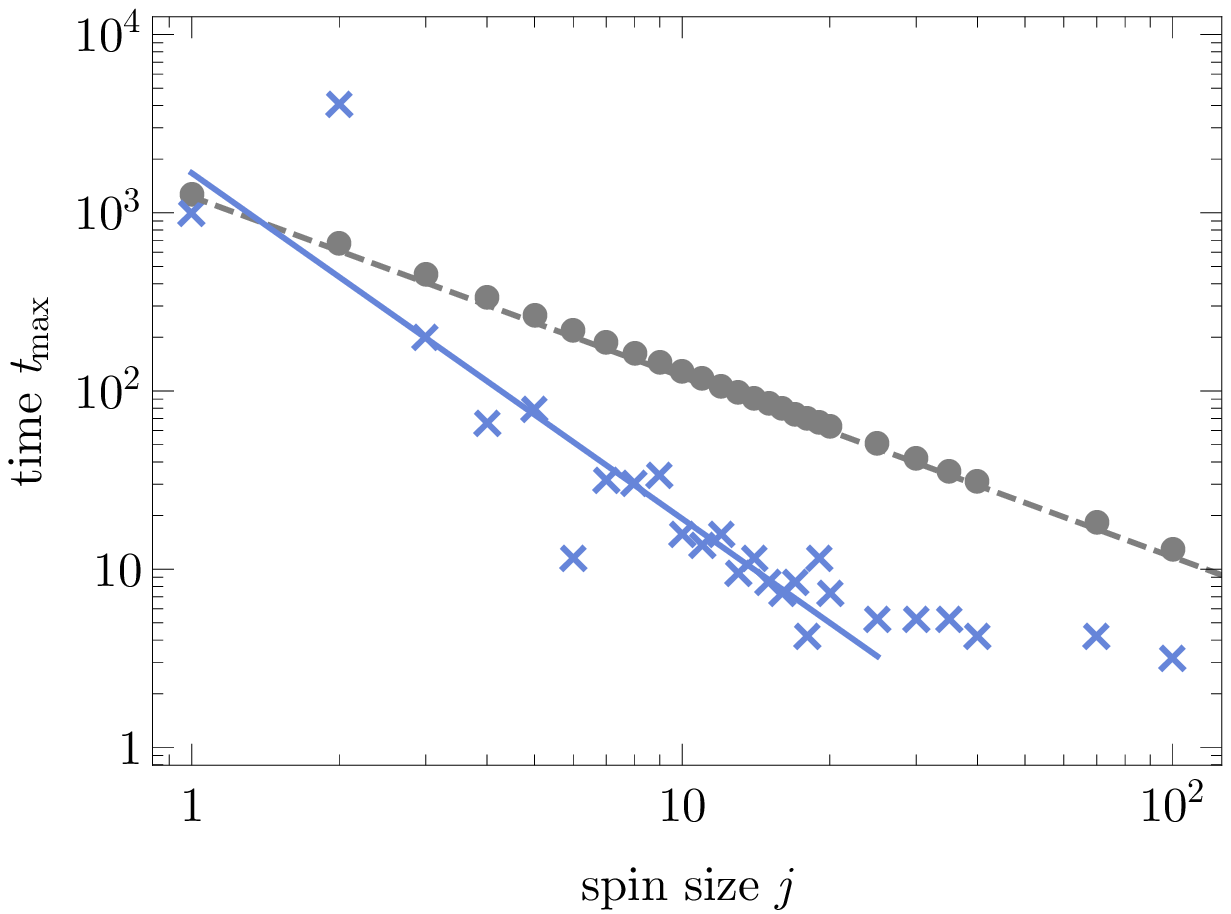}\\
		{\bf (c)}\hspace{8.75cm}{ \bf (d)}\\
	\caption{Time evolution of the quantum Fisher information (QFI) for the dissipative kicked top. 
		In all subplots we look at an initial coherent state at 
		$(\theta,\phi)=(\pi/2,\pi/2)$.
		{\bf(a)} 
		Exemplary behavior as function of time for a
		dissipative top (DT, dashed gray line) and a  
		chaotic (kicking strength $k=30$) dissipative kicked top (DKT, blue crosses) for spin size $j=40$, damping constant $\gamma=0.5\times 10^{-3}$. 
		{\bf(b)} 
		High plateau values of the QFI are achieved for $j=2$ and $k=30$ 
		for various values of the damping constant ($\gamma=0.5\times 10^{-2}$ 
		for the purple line, $\gamma=1.58\times 10^{-3}$ for the orange line, 
		and $\gamma=0.5\times 10^{-4}$ for the blue line), while dashed 
		lines represent corresponding QFI with $k=0$.
		{\bf(c,d)} 
		Time $t_\text{max}$ at which the QFI of the DT 
		(gray dots) and DKT (blue crosses)
		reaches its maximum as function of $\gamma$ and $j$, 
		respectively.  The fits for
		the DT (gray line) exhibit a slope of 
		$-0.97$ and $ -1.01$, and the fits
		for the DKT (blue, dashed) a slope of $-0.95$ and
		$ -1.94$ for scaling with $\gamma$ and $j$,
		respectively. Parameters $j=15$ in {\bf(c)} and
		$\gamma=0.5\times 10^{-3}$ in {\bf (d)}. The 
		plotted range of the fits corresponds to the fitted data.}
	\label{fig:QFI:first:comp}
	\label{fig:high:plateau}
\end{figure}
We find that initial states in the chaotic sea perform best for small
times while for larger times ($t\gtrsim 300$ for $j=4000$) edge states
perform best. Note that this is numerically confirmed up to very
high QFI values ($>10^{14}$). The superiority of edge states holds for
$j\gtrsim 10$ for large times ($t\gtrsim 10^3$, see panel \textbf{(a)} in
Fig.~\ref{fig:KTmix_scaling_compact}). Large values of $j$ allow one to localize
states essentially within a stability island.  
A coherent state localized within
 a non-equatorial stability island
shows a quadratic $t$-scaling analog to the regular top. 
For a state $\ket{\psi_\text{eq}}$
localized around a point within an equatorial island of stability the
QFI drastically decays with increasing $j$ (brown triangles). 
The scaling with $t$ for $j=4000$ reveals that QFI does not increase
with $t$ in this case, it 
freezes. One can understand the phenomenon as arising from a freeze of
fidelity due to a vanishing time averaged 
perturbation \cite{Gorin06review,sankaranarayanan2003recurrence}:
the dynamics restricts the states to the
equatorial stability island with time average
$\overline{\bra{\psi_\text{eq}}(U_\alpha^t)^\dagger(k){J_z}U_\alpha^t(k)\ket{\psi_\text{eq}}}=0$.  
This can be verified  numerically, and contrasted 
with the dynamics when
initial states are localized in the chaotic sea or
on a non-equatorial island. 

\subsection*{Results for the dissipative kicked top}
For any quantum-enhanced measurement, it is important to assess
the influence of dissipation and decoherence. We first study
superradiant damping
\cite{Dicke54,Bonifacio71a,Haake73,Gross82,kaluzny1983observation}
as this 
enables a proof-of-principle demonstration with
an analytically accessible propagator for the
master equation 
with spins up to
$j\simeq200$ and correspondingly large gains. Then, in the next
subsection, we show by detailed and realistic modeling including
all the relevant decoherence mechanisms that sensitivity 
of existing state-of-the-art alkali-vapor-based spin-precession 
magnetometers 
in the spin-exchange-relaxation-free 
(SERF) regime can be enhanced by non-linear kicks.

At sufficiently low temperatures ($k_B
T\ll \hbar\omega$, where $\hbar\omega$ is the level spacing between
adjacent states $\ket{jm}$) superradiance is described by the
Markovian master equation for the spin-density matrix $\rho(t)$ with
continuous time,
\begin{equation}\label{eq:master}
\frac{ d}{ dt}\rho(t)=\gamma ([J_-,\rho(t)
J_+]+[J_-\rho(t),J_+])\equiv 
\Lambda\rho(t)\,, 
\end{equation}
where $J_\pm\equiv J_x\pm iJ_y$, with the 
commutator $[A,B]=AB-BA$,  and $\gamma$ is the dissipation rate, 
 with the formal solution 
 $\rho(t)=\exp(\Lambda t)\rho(0)\equiv
 D(t)\rho(0)$. The full evolution is 
governed by
 $d\rho(t)/dt =\Lambda \rho(t)-i\hbar [H_\textrm{KT}(t),\rho(t)]$.
Dissipation and precession about 
the $z$-axis commute, 
$\Lambda(J_z\rho J_z)=J_z(\Lambda\rho) J_z$.
 $\Lambda$ can therefore act permanently, leading to the propagator $P$ of
$\rho$ from discrete time $t$ to $t+\tau$
for the dissipative kicked top ($\textrm{DKT}$) \cite{haake2013quantum,Braun01B}
\begin{equation}\label{eq:DKT:map}
\rho(t+\tau)=P\rho(t)=U_{\alpha}(k)\left(D(\tau)\rho(t)\right)U_{\alpha}^\dagger(k).
\end{equation}
For the sake of simplicity we again set the period $\tau=1$, and $t$
is taken again as discrete time in units of $\tau$.  Then,
$\gamma\tau\equiv\gamma$  
controls the effective dissipation between
two unitary propagations. Classically, the $\textrm{DKT}$ shows
a strange attractor in
phase space with a fractal dimension that reduces from $d=2$ at
$\gamma=0$ to $d=0$ for large $\gamma$, when the attractor shrinks to
a point attractor and migrates towards the 
ground state $\ket{j,-j}$ \cite{Braun01B}. 
Quantum mechanically, one finds a Wigner function with
support on a smeared out version of the strange attractor that
describes a non-equilibrium steady state reached after many
iterations. Such a non-trivial state is only possible through the
periodic addition of energy due to the kicking. Because of the 
filigrane structure of the strange attractor, one might hope
for relatively large QFI, whereas without
kicking the system would decay to the ground state, where the QFI
vanishes. Creation of steady non-equilibrium states may therefore
offer a way out of the decoherence problem in quantum metrology, 
see also section V.C in Ref.~\cite{braun2017without} for similar ideas.

Vanishing kicking strength, i.e.~the dissipative top 
(DT)  obtained from the $\textrm{DKT}$  by setting $k=0$,  
will serve again as benchmark. While in the dissipation-free
regime we took the top's QFI and with it its SQL-scaling
($\propto jt^2$) as reference, SQL-scaling no longer represents a
proper benchmark, because damping typically corrupts QFI with
increasing time. To illustrate the typical behavior of QFI we
exemplarily choose certain spin sizes $j$ and damping
constants $\gamma$ here and in the following, such
as $j=40$ and $\gamma=0.5\times10^{-3}$ in
Fig.~\ref{fig:QFI:first:comp} \textbf{(a)}, while
computational limitations restrict us to $j\lesssim200$. 

Figure \ref{fig:QFI:first:comp} shows the typical
overall behavior of the QFI of the $\textrm{DT}$ and $\textrm{DKT}$ 
as function of time: After a 
steep initial rise $\propto t^2$, the QFI reaches a maximum whose
value is the larger 
the smaller the dissipation.  Then the QFI decays again, dropping to
zero for the $\textrm{DT}$, and a plateau value for the $\textrm{DKT}$. 
The time at which
the maximum value is reached decays roughly as $1/(j\gamma)$ for 
the $\textrm{DT}$, and as $1/j^{0.95}$ and $1/\gamma^{1.94}$
for the $\textrm{DKT}$. The plateau itself is in general relatively small for the
limited values of $j$ that could be investigated numerically, but it
should be kept in mind that i.) for the DT the plateau does not even
exist (QFI always decays to zero for large time, as dissipation drives
the system to the ground state $\ket{j,-j}$ which is an eigenstate of
$J_z$ and hence insensitive to precession); and ii.), there are
exceptionally large plateau values even for small $j$, see e.g.~the
case of $j=2$ in Fig.~\ref{fig:high:plateau} \textbf{(b)}.  
There, for $\gamma=1.58\times 10^{-3}$, the plateau value is 
larger
by a factor $2.35$ than the DT's QFI optimized over all 
initial coherent states for all times. 
Note that  since $\Lambda (J_z\rho J_z)=J_z
(\Lambda \rho) J_z$, for the DT
an initial precession about the $z$-axis that is part of the state
preparation can 
be moved to the end of the evolution and does not influence the
QFI of the DT. Optimizing over the initial coherent
state can thus be restricted to
optimizing  over $\theta$. \\

When considering dynamics, it is natural also to include time
as a resource. Indeed, experimental sensitivities are normally given
as uncertainties per square root of
Hertz:  longer (classical) averaging
reduces the uncertainty as $1/\sqrt{T_\text{av}}$ with averaging
time $t=T_\text{av}$. For fair comparisons, one multiplies
the achieved uncertainty with $\sqrt{T_\text{av}}$.  Correspondingly,
we now compare rescaled QFI and Fisher information, 
namely $I_\alpha^{(t)}\equiv
I_\alpha/T_\text{av}$, $I_{\textrm{Fisher},\alpha}^{(t)}\equiv
I_{\textrm{Fisher},\alpha}/T_\textrm{av}$. 
A protocol which reaches
a given level of QFI more rapidly has then an advantage, 
and best precision corresponds to the maximum 
rescaled QFI or Fisher information, $\hat{I}_\alpha^{(t)}\equiv\max_t
I^{(t)}_\alpha$ or $\hat{I}_{\textrm{Fisher},\alpha}^{(t)}\equiv\max_t
I^{(t)}_{\textrm{Fisher},\alpha}$. 

\begin{figure}
		\includegraphics[width=9cm]{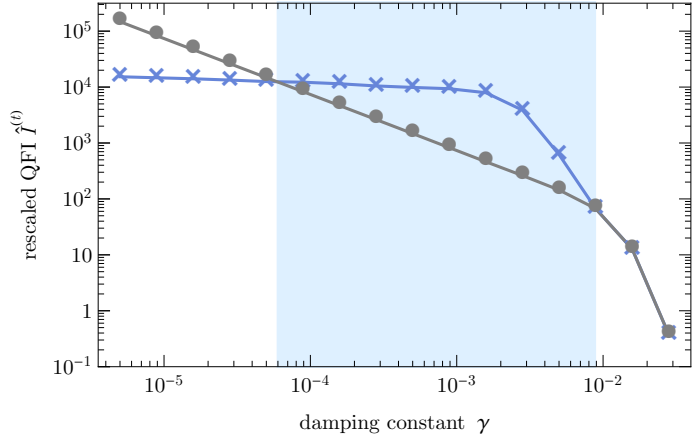}
	\caption{Enhancement in measurement precision through kicking (blue crosses) is found over a 
		broad range of damping strengths. Comparison of the maximal rescaled quantum 
		Fisher information	$\hat{I}_\alpha^{(t)}$ for the dissipative top 
		(gray dots) and the dissipative kicked top (blue
		crosses); kicking strength $k=30$ and spin size $j=100$. In both cases,
		$\hat{I}_\alpha^{(t)}$ was optimized over the location of
		initial coherent states. The blue-shaded area marks the
		range $1.2\lesssim2\gamma J^2\lesssim180$ where the 
		reference is outperformed.}
	\label{fig:kappacomp1}
\end{figure}
Figure \ref{fig:kappacomp1} shows that in a broad range of 
dampings that are sufficiently strong for the QFI to decay early,
the maximum rescaled QFI of the $\textrm{DKT}$ beats that quantity of the
DT by up to an 
order of magnitude. Both quantities were optimized over the location
of the initial coherent states.

\begin{figure}
		\includegraphics[width=9cm]{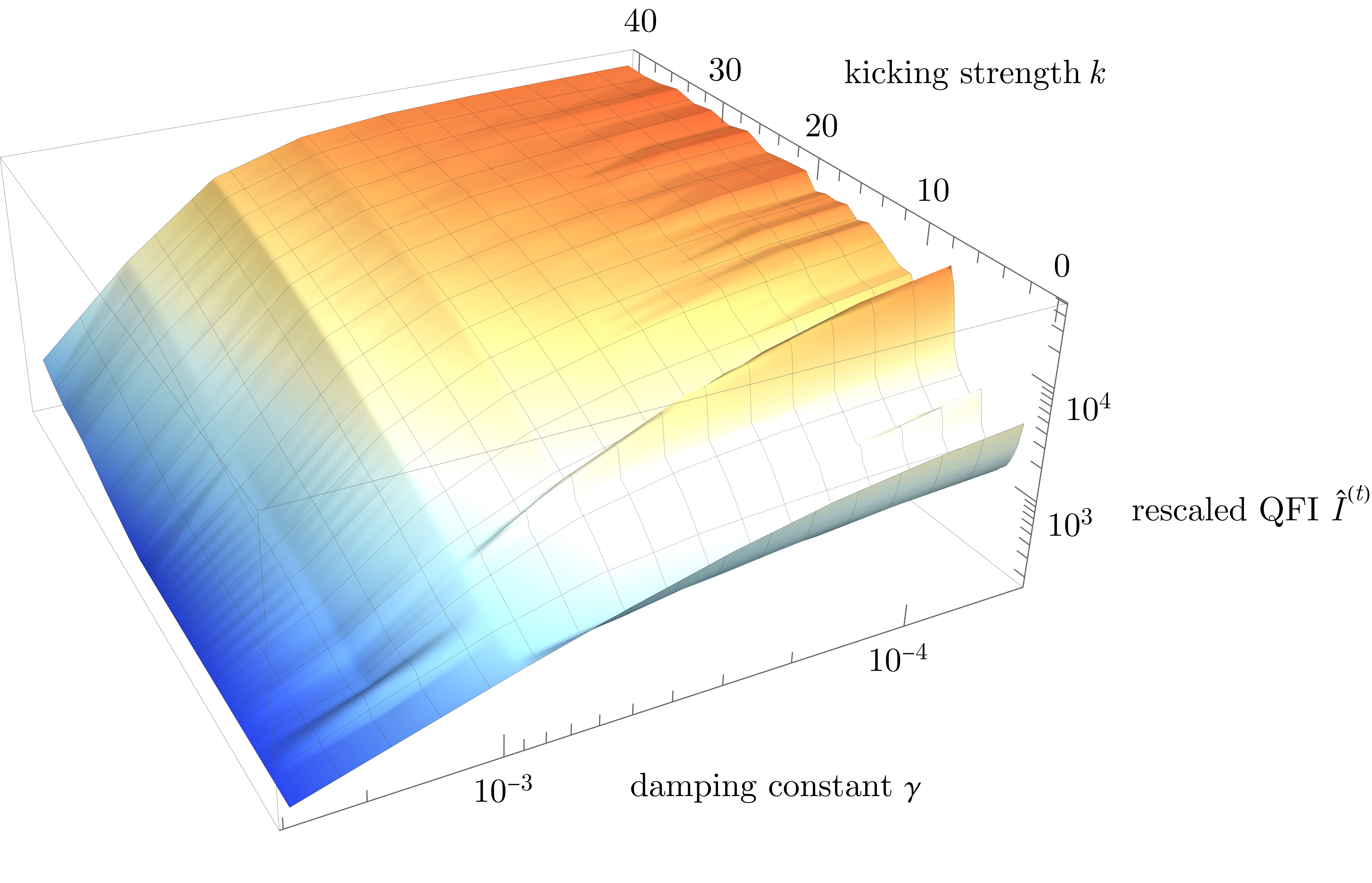}
		\includegraphics[width=9cm]{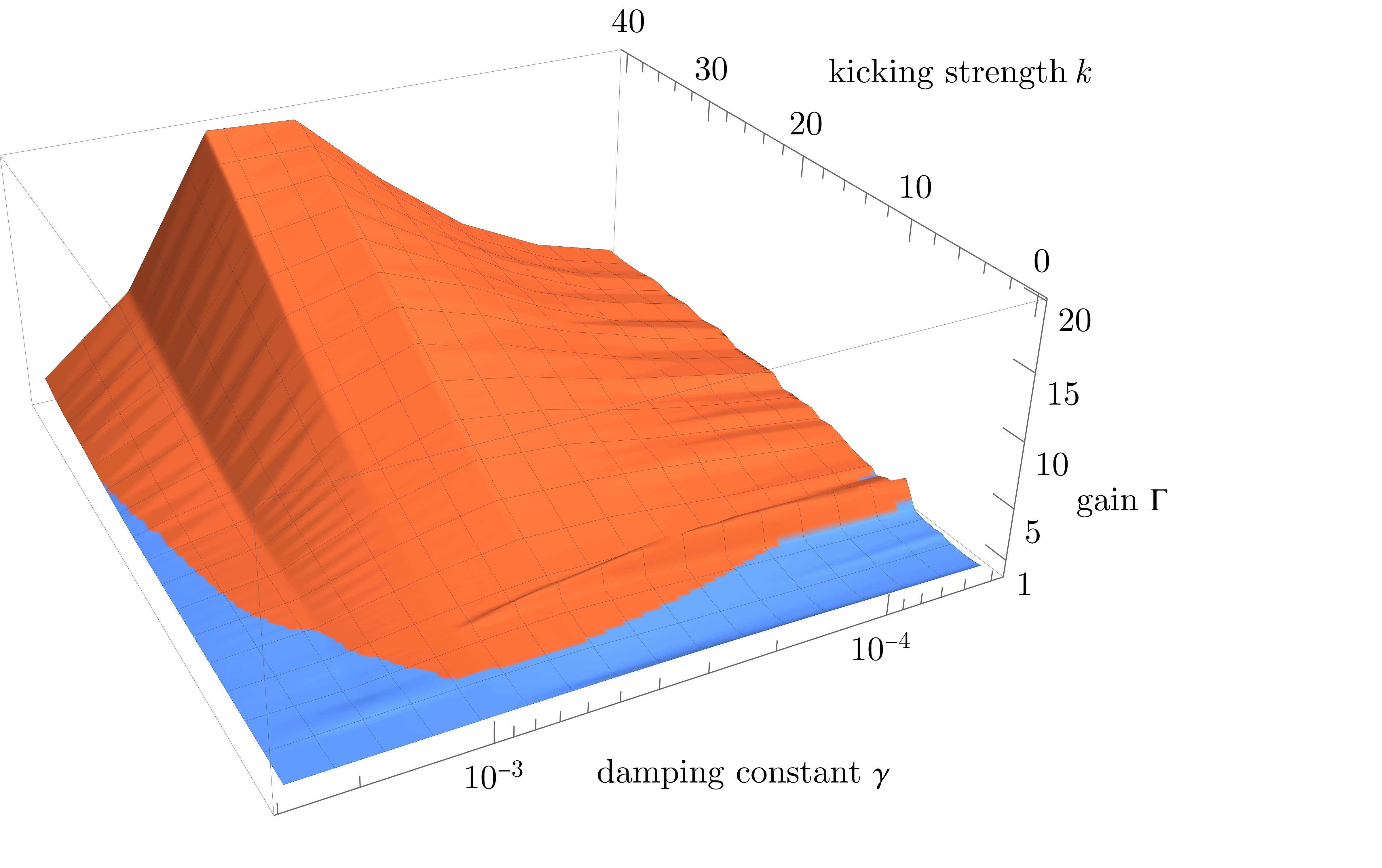}\\
		{\bf(a)}\hspace{9cm}{\bf(b)}
	\caption{Rescaled quantum Fisher information (QFI) and gains in measurement precision. 
		Rescaled QFI 
		of the dissipative kicked top
		$\hat{I}^{(t)}_{\alpha,\textrm{DKT}}(\theta=\pi/2,\phi=\pi/2)$ in panel \textbf{(a)}
		and gain $\Gamma=\frac{\hat{I}_{\alpha,\textrm{DKT}}^{(t)}(\theta=\pi/2,\phi=\pi/2)}{\hat{I}_{\alpha,\text{DT}}^{(t)}}$ in panel \textbf{(b)}
		as function of damping constant $\gamma$ and
		kicking strength $k$ at $j=200$. Note that in this case $\hat{I}^{(t)}_{\alpha}$ is 
		optimized over initial states only for the dissipative top.}   
	\label{fig:kappacomp}
\end{figure}
Figure \ref{fig:kappacomp} shows 
$\hat{I}_\alpha^{(t)}$ and the gain in that quantity compared to the
non-kicked case as function of both the damping and the kicking
strength. One sees that in the intermediate damping regime ($\gamma\simeq
10^{-3}$) 
the gain increases with kicking strength, i.e.~increasingly 
chaotic dynamics. \\

\begin{figure}[h!]
	\includegraphics[width=9cm]{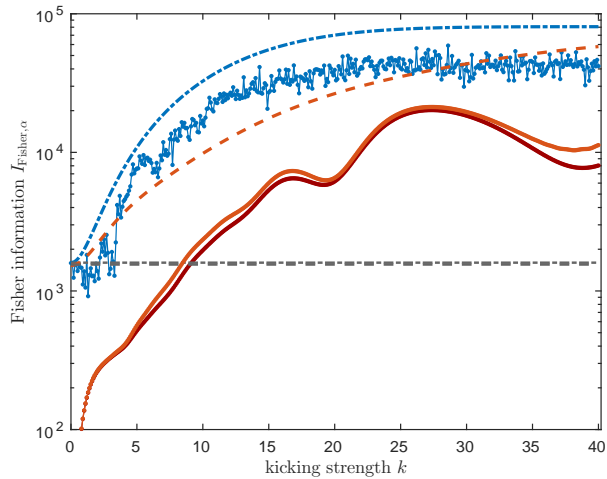}
	\caption{Performance of an exemplary spin-component
		measurement with superradiance damping. Fisher information $I_{\text{Fisher},\alpha}$
		related to measuring the $y$-component of the spin, $J_y$, after $t=2$
		time steps without dissipation (blue line) as well as in the
		presence of dissipation (bright red line, damping constant $\gamma=0.5\times10^{-3}$), 
		upper bounded by corresponding quantum Fisher informations (blue dash-dotted and red dashed 
		lines), for spin size $j=200$, and an initial state at $(\theta,\phi)=(\pi/2,\pi/2)$. 
		For the dissipative case Fisher information with a $5\%$ uncertainty 
		in the kicking strength is given by the dark red line slightly 
		below the bright red line.  Horizontal gray 
		lines represent benchmarks (kicking strength $k=0$) for $t=2$ optimized 
		over initial states (dashed with dissipation and dash-dotted without; lines are almost on top of each other). 
		In both cases the benchmark is clearly outperformed 
		for $k\gtrsim 3.5$ and $k\gtrsim 10.6$ without 
		and with dissipation, respectively.}
	\label{fig:Jy:dissi}
\end{figure}
For exploiting the enhanced
sensitivity shown to exist through the large QFI, one needs also to
specify the actual measurement of the probe.  In principle the QCRB
formalism allows one to identify the optimal POVM measurement if the
parameter is known, but these may not always be realistic. In
Fig.~\ref{fig:Jy:dissi} we investigate $J_y$ as a  feasible example for a
measurement for a spin size $j=200$ after $t=2$ time steps. We find
that there exists a broad range of kicking strengths where the
reference (state-optimized but $k=0$) is outperformed in both cases,
with and without dissipation. 
In a realistic experiment,
control parameters such as the kicking strength are subjected to
variations. A $5\%$ variance in $k$, which was reported in
Ref.~\cite{chaudhury2009quantum}, reduces the Fisher information
only marginally and does not challenge the advantage of
kicking. This can be calculated by rewriting the probability that
enters in the Fisher information in eq.~(\ref{eq:Fisher}) according to
the law of total probability,  $p_\alpha(\xi)=\int dk\,
p(k)p_\alpha(\xi|k)$ where $p(k)$ is an assumed Gaussian
distribution of $k$ values with $5\%$ variance and
$p_\alpha(\xi|k)=\tr[\Pi_\xi\rho_\alpha(k)]$ with $\Pi_\xi$ a POVM
element and $\rho_\alpha(k)$ the state for a given $k$ value. The
advantage from kicking remains when investigating a rescaled and
time-optimized Fisher information (not shown in
Fig.~\ref{fig:Jy:dissi}).
  
\subsection*{Improving a SERF magnetometer}
We finally show that quantum-chaotically enhanced sensitivity can
be achieved in state-of-the-art magnetometers by investigating a
rather realistic and detailed  model of an alkali-vapor-based
spin-precession magnetometer acting in the
SERF regime.  SERF magentometers 
count amongst the most sensitive magnetometers for detecting small
quasi-static magnetic fields
\cite{allred2002high,Kominis03,savukov2005effects,budker2007optical,sheng2013subfemtotesla}. We
consider a cesium-vapor magnetometer at room temperature in the SERF regime similar to experiments with rubidium in
Ref.~\cite{balabas2010polarized}. Kicks on the single cesium-atom
spins can be realized as in Ref.~\cite{chaudhury2009quantum} by exploiting
the spin-dependent rank-2 (ac 
Stark) light-shift generated with the help of an off-resonant laser
pulse. Typical SERF magnetometers working at higher temperatures with high buffer-gas pressures exhibit an unresolved excited state hyperfine splitting due to pressure broadening, which makes kicks based on rank-2 light-shifts ineffective. Dynamics are modeled in the electronic ground state
$6^2\text{S}_{1/2}$ of $^{133}$Cs that splits into total spins of 
$f=3$ and $f=4$, where kicks predominantly act on the $f=3$ manifold.

The model is quite different from the foregoing superradiance
model because of a different decoherence mechanism originating from
collisions of Cs atoms in the vapor cell: We include spin-exchange and spin-destruction relaxation, as well as additional decoherence
induced by the optical implementation of the kicks. 
With this
implementation of kicks one is confined to a small spin size $f=3$ of single atoms,
such that the large improvements in sensitivity found for the
large spins discussed above cannot be expected. Nevertheless, we
still find a clear gain in the sensitivity and an improved robustness to decoherence due to kicking.  Details of the
model described with a master equation \cite{appelt1998theory,deutsch2010quantum} can
be found in the Supplementary Note 2.

Spins of cesium atoms are initially pumped into a state
spin-polarized in $z$-direction orthogonal to the magnetic field
$\textbf{B}=B\hat{\textbf{y}}$ in $y$-direction, whose strength $B$
is the
parameter $\alpha$ to be measured. We let spins precess in
the magnetic field, and,  by incorporating small kicks 
about the $x$-axis, we find an improvement over the reference
(without kicks) in terms of rescaled QFI and the
precision based on  the measurement of the electron-spin component $S_z$ orthogonal to the magnetic field. The
best possible measurement precision $\Delta B$ in units of 
T$/\sqrt{\textrm{Hz}}$ per $1\,\textrm{cm}^3$ vapor volume is
$\Delta B=1/\sqrt{nI_B^{(t)}}$ where $n\simeq 2\times 10^{10}$ is the number
of cesium atoms in $1\,\textrm{cm}^3$. For a specific measurement,
$I_B^{(t)}$ must be replaced by the corresponding rescaled Fisher
information $I_\textrm{Fisher, B}^{(t)}$. We compare the models with
and without kick directly on the basis of the Fisher information 
rather than modeling in addition the
specific optical implementation and the 
corresponding  noise
of the measurement of $S_z$. Neglecting this additional
read-out-specific noise leads to slightly better precision  
bounds than given in the literature, but does not distort the
comparison.   

The magnetic field was set to $B=4\times10^{-14}\,\textrm{T}$
in $y$-direction, such that the condition for the SERF regime is fulfilled, i.e.~the Larmor frequency is much smaller than the spin-exchange rate, and the period is set to $\tau=1\,$ms. 
Since kicks induce decoherence in the atomic spin system, 
we have to choose a very small effective kicking strength of 
$k\simeq 6.5\times 10^{-4}$ for the kicks around the $x$-axis (with respect to the $f=3$ ground-state manifold), 
generated with an off-resonant $2\,\mu$s light pulse with 
intensity $I_\text{kick}=0.1\,$mW/cm$^2$ linearly polarized in $x$-direction, 
to find an advantage over the reference.

\begin{figure}[h]
		\centering
		\includegraphics[width=9cm]{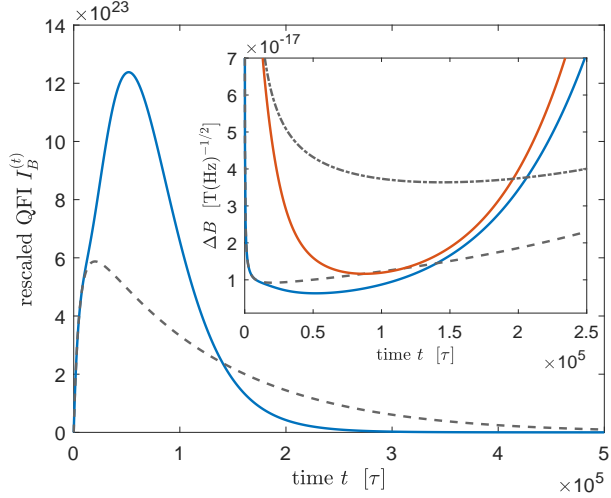}
	\caption{Performance of a magnetic-field measurement with a kicked atomic-vapor magnetometer. While the gray dashed line
		shows the rescaled quantum Fisher information $I^{(t)}_B$ for measuring the magnetic field $B$ with an spin-exchange-relaxation-free magnetometer, the blue line is obtained
		by adding a short optical kick at the end of each period $\tau$. The inset shows
		precision $\Delta B$ in units of T per $\sqrt{\text{Hz}}$ for an
		optimal measurement (gray dashed line and blue line without and with kicks, respectively) and for measuring the $z$-component of the electronic spin $S_z$ (grey dash-dotted and red line without and with kicks, respectively).}\label{fig:SERF}
\end{figure}
The 
example of Fig.~\ref{fig:SERF} shows about $31\%$ improvement 
in measurement precision $\Delta B$ for an optimal measurement (QFI,
upper right inset) and $68\%$ improvement in a comparison of 
$S_z$ measurements (inset), which is impressive in view of the 
small system size.The achievable measurement precision of the kicked dynamics exhibits an improved  robostness to decoherence: rescaled QFI for the kicked dynamics continues to increase and sets itself apart from the reference around the coherence time associated with spin-destruction relaxation.
The laser light for these pulses can be provided by the laser
used for the read out, 
which is typically performed with an
off-resonant  laser. A further improvement in precision is expected
from additionally measuring kick pulses for readout or by applying
kicks not only to the $f=3$ but also to the $f=4$ ground-state 
manifold of $^{133}$Cs. Further, it might be possible to
dramatically increase the relevant spin-size by
applying the kicks to the 
joint spin of the cesium atoms, for instance, through a double-pass
Faraday effect \cite{takeuchi2005spin}.
    
\section*{Discussion}
Rendering the dynamics of quantum sensors chaotic allows one to
harvest a quantum enhancement for quantum metrology without
having to rely on the preparation or stabilization 
of highly entangled states.  
 Our results imply that existing magnetic
field sensors \cite{budker2007optical,ledbetter2008spin} based
on the precession of a spin can be 
rendered more sensitive by disrupting the time-evolution by
non-linear kicks. The enhancement persists in rather broad parameter
regimes even when including the effects of dissipation and
decoherence. Besides a thorough investigation of superradiance damping
over large ranges of parameters, we studied a cesium-vapor-based atomic
magnetomter in the SERF regime based on a detailed and realistic 
model
\cite{allred2002high,Kominis03,savukov2005effects,budker2007optical,balabas2010polarized}. 
Although the implementation of the non-linearity via a rank-2 light shift
introduces additional decoherence and despite the rather small atomic spin size
$f\leq 4$, a considerable improvement in measurement sensitivity is
found ($68\%$ for a read-out scheme based on the measurement of the 
electronic spin-component $S_z$).
The required non-linearity that can be
modulated as function of time has been demonstrated  experimentally in
\cite{chaudhury2009quantum} in cold cesium vapor.  

Even higher gains in
sensitivity are to be expected if an effective interaction can be
created between the atoms, as this opens access to larger values of
total spin size for the kicks. This may be achieved e.g.~via a
cavity as suggested for pseudo-spins in \cite{agarwal1997atomic}, or
the interaction with a propagating light field as demonstrated
experimentally in \cite{takeuchi2005spin,Julsgaard01} with about
$10^{12}$ cesium 
atoms. More generally, our 
scheme will profit from the accumulated
knowledge of spin-squeezing, which is 
also based on the creation of an
effective interaction between atoms. Finally, we expect that improved
precision can be found in 
other quantum 
sensors that can be rendered chaotic as well, as the
underlying sensitivity 
to change of parameters is a basic property of quantum-chaotic
systems. 

\section*{Methods}
\subsection*{QFI for kicked time-evolution of a pure state}
The QFI in the chaotic regime with large system dimension $2J$
and times larger than the Ehrenfest time, $t>t_\text{E}$, is given in
linear-response theory by an auto-correlation function 
$C(t)\equiv
\braket{\tilde{V}(t)\tilde{V}(0)}-\braket{\tilde{V}(t)}\braket{\tilde{V}(0)}$
of the perturbation of the Hamiltonian in the
interaction picture, 
$H_{\alpha+\epsilon}(t)=H_\alpha(t)+\epsilon V(t)$, $\tilde{V}(t)=U_\alpha(-t)V(t)U_\alpha(t)$:  
\begin{equation}\label{eq:qfi:chaos:ehrenfest}
I_\alpha(t)=4\left(tC(0)+2\sum_{t'=0}^{t-1} (t-t') C(t')\right).
\end{equation}
In our case, the perturbation $V(t)=J_z$ is proportional to the 
parameter-encoding precession Hamiltonian, and the first summand in
eq.~\eqref{eq:qfi:chaos:ehrenfest} can be calculated for 
an initial coherent state,
\begin{equation}\label{eq:qfi:chaos:ehrenfest:1}
C(0)=\frac{1}{3}j(j+1),
\end{equation}
giving a $tj^2$-scaling starting from $t_\text{E}$. Due to the finite
Hilbert-space dimension of the kicked top, the auto-correlation
function decays for large times to a finite value $\bar{C}$, leading to a term
quadratic in $t$ from the sum in
eq.~\eqref{eq:qfi:chaos:ehrenfest} that simplifies to 
$I_\alpha=4\overline{C}t^2$ for $t\gg t_H$. 
If one rescales $J_z\rightarrow J_z/J$ such that it has a well defined classical limit, random matrix theory allows one to
estimate the average value of $C(t)$ for large times:  $\bar{C}=2Js\sigma_\text{cl}$, and 
$\sigma_\text{cl}$ is a transport coefficient that can be calculated
numerically \cite{Gorin06review}. This yields
eq.~(\ref{eq:fidelity:chaos:Heisenberg}). 

\subsection*{Lyapunov exponent} 
A data point located at $(Z,\phi)$ for the Lyapunov exponent in Fig.~2
\textbf{(c)} was obtained numerically by 
averaging over $100$ initial conditions equally distributed within a 
circular area of size $1/j$ (corresponding to the coherent state) 
centered around  $(Z,\phi)$. \\

\subsection*{Data availability}
Relevant data is also available from the authors upon request.

\section*{End notes}
\subsection*{Acknowledgements}
This work was supported by the Deutsche Forschungsgemeinschaft (DFG), Grant No. BR 5221/1-1. Numerical calculations were performed in part with resources supported by the Zentrum f{\"u}r Datenverarbeitung of the University of T{\"u}bingen.
\subsection*{Author Contribution}
D.~B.~initiated the idea and L.~F.~made the calculations and  numerical simulations. Both Authors contributed to the interpretation of data and the writing of the manuscript.
\subsection*{Competing Interests}
The authors declare no competing interests.

\newpage
\section*{Supplementary Information}
\subsection*{Supplementary Note 1. Realization of the kicked top}
The kicked top (KT) has been realized experimentally by Chaudhury et al.~\cite{chaudhury2009quantumsup}
using the atomic spin of a $^{133}\textrm{Cs}$ atom in the $f=3$
hyperfine ground state. Linear precession
of the spin was implemented through magnetic pulses, and the 
torsion
through an off-resonant laser field that exploited a spin-dependent rank-2 (ac Stark) light shift. \\

An implementation of  the KT using microwave
superradiance was proposed by Haake \cite{haake2000cansup}: The top is
represented by the collective pseudo-spin of $N$ two-level 
atoms coupled with the same coupling constant to a single mode of an
electromagnetic field in a cavity, with a
controlled detuning between mode and atomic frequencies. 
Large detuning compared to the Rabi frequency
$\Omega=g\sqrt{N}$ with coupling strength $g$ allows one to adiabatically eliminate the cavity
mode and leads to an effective interaction of the type $J_z^2$
\cite{agarwal1997atomicsup} (replacing our $J_y^2$), 
while superradiant damping as described by 
the  master equation (13) in the main text
for the reduced density operator of the atoms can
still prevail \cite{haake2000cansup}. Finally, a linear rotation about
the $x$-axis can be achieved through resonant microwave pulses,
replacing the linear precession about the $z$-axis of the KT.
The parameter $\alpha$ is now proportional to the Rabi frequency of
the microwave pulse.

\subsection*{Supplementary Note 2. Spin-exchange-relaxation-free Cs-vapor magnetometer}\label{meth:SERF}
Adapting standard notation in atomic physics, atomic spin operators
will be denoted in the following by $\textbf{F}=(F_x,F_y,F_z)$ with
spin size $f$, $F_z\ket{fm}=m\ket{fm}$, and total electronic angular momentum
$\textbf{J}=\textbf{L}+\textbf{S}$ 
with quantum number $j$, composed of orbital angular momentum
$\textbf{L}$ and electron spin $\textbf{S}$.  
We model a room-temperature spin-exchange-relaxation-free (SERF) Cs-vapor magnetometer
similar to the experiments with Rb-vapor of Balabas et al.~\cite{balabas2010polarizedsup}.  The Cs spin sensitive to the magnetic field
$B$ is composed of a nuclear spin $K=7/2$ and one valence electron
with an electronic spin $s=1/2$ which splits the ground state
$6^2\text{S}_{1/2}$ into two energy levels with total spin $f_1=3$ and
$f_2=4$. This results in an effective Hilbert space of dimension
$2(2K+1)=16$ for our model of a kicked SERF magnetometer. 

The dominant damping mechanisms are related to collisions of cesium
atoms with each other and with the walls of the vapor cell.

In the SERF regime the spin-exchange rate is much greater than the 
rate of Larmor precession, typically realized by very small magnetic fields, 
a high alkali-atom density ($10^{13}$ atoms per cm$^3$), high buffer-gas pressure, and heating of the vapor cell. Then, spin-exchange relaxation is so strong, that the  population 
of hyperfine ground levels ($f_i=3,4$) is well described by a spin-temperature 
distribution. Here, we model a SERF magnetmeter with a lower alkali-atom density of $2\times 10^{10}$ atoms per cm$^3$ without buffer gas at room temperature $T=294.13\,$K. Modern alkene-based vapor-cell coatings support up to $10^6$ collisions before atoms become depolarized \cite{balabas2010polarizedsup}. For a spherical vapor-cell with a $1.5\,$cm radius it follows that collisions with walls limit the lifetime of spin polarization to $T_\textrm{wall}\simeq 92\,$s. Since the effect of collisions among Cs atoms leads to stronger depolarization we neglect collisions with the walls in our model. 

While the typical treatment proceeds by eliminating 
the nuclear-spin component we are interested in a dynamics that 
exploits the larger Hilbert space of the Cs spin. Therefore the 
evolution of the spin density matrix $\rho$ is described by a master 
equation that includes damping originating from collisions of Cs 
atoms \cite{appelt1998theorysup} and an interaction with an off-resonant 
light field in the low-saturation limit \cite{deutsch2010quantumsup} 
modeling the kicks:
\begin{align}\label{eq:SERFmaster}
\frac{d\rho}{dt}&=R_\textrm{se}\left[\varphi(1+4\braket{\textbf{S}}\cdot\textbf{S})-\rho\right]+R_{\text{sd}}\left[\varphi-\rho\right] 
+a_\textrm{hf}\frac{[\textbf{K}\cdot\textbf{S},\rho]}{i\hbar}+\frac{H_A^{\text{eff}}\rho-\rho H_A^{\text{eff}\dagger}}{i\hbar} 
\notag\\&\quad +\gamma_\text{nat}\sum_{q=-1}^1\left(\sum_{f,f_1}W_q^{ff_1}\rho_{f_1f_1}\left(W_q^{ff_1}\right)^\dagger+\sum_{f_1\neq f_2}W_q^{f_2f_2}\rho_{f_2f_1}\left(W_q^{f_1f_1}\right)^\dagger\right)
\end{align}
The first two summands describe spin-exchange 
relaxation and spin-destruction relaxation, respectively, where
$R_\text{se}$ 
denotes the spin-exchange rate and  $R_\text{sd}$ the spin-destruction 
rate, and $\varphi=\rho/4+\textbf{S}\cdot\rho\textbf{S}$ is called the purely 
nuclear part of the density matrix, where the electron-spin operator $\textbf{S}$ 
only acts on the electron-spin component with expectation value  
$\braket{\textbf{S}}=\tr[\textbf{S}\rho]$. 
The third summand is the hyperfine coupling of nuclear spin \textbf{K} 
and electronic spin \textbf{S} with hyperfine structure constant $a_\text{hf}$, 
and the fourth summand drives the dynamic with an effective
non-hermitian Hamiltonian on both ground-state hyperfine manifolds $H_A^\text{eff}=H_{A,f=3}^\text{eff}+H_{A,f=4}^\text{eff}$, with
\begin{equation}
H_{A,f}^\text{eff}=\hbar\Omega_\text{Lar}F_y+\sum_{f'} \frac{\hbar\Omega^2 C_{j'f'f}^{(2)}}{4(\Delta_{ff'}+i\gamma_\text{nat}/2)}\left|\boldsymbol{\epsilon_\text{L}}\cdot\textbf{F}\right|^2
\end{equation}
that includes Larmor precession with frequency $\Omega_\text{Lar}=g_f\mu_BB/\hbar$ 
(with the Land\'e g-factor $g_f$ and the Bohr magneton $\mu_B$) of the 
atomic spin in the external magnetic field $\textbf{B}=B\hat{\textbf{y}}$, 
and the rank-2 light-shift induced by a light pulse that is linearly 
polarized with unit polarization vector $\boldsymbol{\epsilon}_\text{L}$ 
of the light field and off-resonant with detuning $\Delta_{ff'}$ from 
the D1-line transition with $f\rightarrow f'$. 
Further, we have the characteristic Rabi frequency
$\Omega=\gamma_\text{nat}\sqrt{I_\text{kick}/(2I_\text{sat}) }$ of the D1 line,
the natural line width $\gamma_\text{nat}$, kick-laser intensity $I_\text{kick}$, saturation intensity
$I_\text{sat}$, and the coefficient 
\begin{equation}
C_{j'f'f}^{(2)}=(-1)^{3f-f'}\frac{\sqrt{30}(2f'+1)}{\sqrt{f(f+1)(2f+1)(2f-1)(2f+3)}} \left\{\begin{matrix}
f & 1 & f'\\ 1 & f & 2 \end{matrix}\right\}\left|o_{1/2 f}^{j'f'}\right|^2,
\end{equation}
where the curly braces denote the Wigner $6j$ symbol and
\begin{equation}\label{eq:ofunc}
o_{jf}^{j'f'}=(-1)^{f'+1+j'+K}\sqrt{(2j'+1)(2f+1)}\left\{\begin{matrix}
f & K & j'\\ j & 1 & f \end{matrix}\right\},
\end{equation}
where total angular momentum of ground and excited levels of the D1 
line are $j=j'=1/2$.
Photon scattering is taken into account by the imaginary shift of 
$\Delta_{ff'}$ in the effective Hamiltonian and by the remaining parts 
of the master equation that correspond to optical pumping which leads 
to cycles of excitation to the 6P$_{1/2}$ manifold and spontaneous 
emission to the ground-electronic manifold 6S$_{1/2}$. When the laser 
is switched off the master equation solely involves the first four 
summands where $H_A^{\text{eff}}$ reduces to the Larmor precession term.

The jump operators are given as
\begin{align}
W_q^{f_bf_a}=\sum_{f'=3}^4\frac{\Omega/2}{\Delta_{f_af'}+i\gamma_\text{nat}/2}\left(\textbf{e}_q^*\cdot\textbf{D}_{f_bf'}\right)\left(\boldsymbol{\epsilon}_\textrm{L}\cdot\textbf{D}_{f_af'}^\dagger\right),
\end{align}
with the spherical basis $\textbf{e}_1=-(\hat{\textbf{x}}+i\hat{\textbf{y}})/\sqrt{2}$,
$\textbf{e}_0=\hat{\textbf{z}}$,  $\textbf e_{-1}=(\hat{\textbf{x}}-i\hat{\textbf{y}})/\sqrt{2}$ in the Cartesian basis $\hat{\textbf{x}}, \hat{\textbf{y}}, \hat{\textbf{z}}$, and the raising operator $\textbf{D}^\dagger_{ff'}=\sum_{q,m,m'}\textbf{e}_q^*o_{jf}^{j'f'}\braket{f'm'|fm;1q}\ket{f'm'}\bra{fm}$ with Clebsch-Gordan coefficients $\braket{f'm'|fm;1q}$ and magnetic quantum numbers $m$, $m'$.

Excited state hyperfine levels are Doppler and pressure broadened, but we neglect pressure broadening which is much smaller than Doppler broadening due to the very low vapor pressure. Doppler broadening is taken into account by numerically averaging the righthand side of the master equation (Supplementary Equation \ref{eq:SERFmaster}) over the Maxwell-Boltzmann distribution of velocities of an alkali atom. This translates into an average over detunings $\Delta_{ff'}$. Since $\Delta\gg\Omega$ must hold within the description of this master equation, we limit averaging over detunings to a $3\sigma$ interval.

By numerically solving the non-linear trace-preserving master equation 
(Supplementary Equation \ref{eq:SERFmaster}) with the Euler method (analog to 
Ref.~\cite{savukov2005effectssup} we take hyperfine coupling into account 
by setting off-diagonal blocks of the density matrix in the coupled 
$\ket{fm}$-basis after each Euler step to zero, because they oscillate 
very quickly with $a_\text{hf}$),  we simulate dynamics similar to the 
dissipative kicked top described above with the difference that kicks are not 
assumed to be arbitrarily short, i.e.~kicks and precession coexist 
during a light pulse. Kicks and corresponding dissipation are factored in 
by applying a superoperator to the state in each Euler step during a kick.

Spin-exchange and spin-destruction rates are estimated to $R_\textrm{se}\simeq 12\,$Hz and $R_\textrm{sd}\simeq0.12\,$Hz from the known cross sections of Cs-Cs collisions, the mean relative thermal velocity of Cs atoms and their density.

For the concrete example of Fig.~7 we calculate with $2\times10^{10}$
Cs atoms per $1\,$cm$^3$ vapor volume, and
kick laser pulses linearly polarized in $x$-direction, 
$\boldsymbol{\epsilon}_\textbf{L}=\textbf{x}$, with intensity 
$I_\text{kick}=0.1\,\text{mW/cm}^2$ and detuning halfway between 
the two components of the D1 line, and $\Delta_{34}\simeq
-584\,\textrm{MHz}$.   
The period is $\tau=1\,\textrm{ms}$ where during the last 
$2\,\mu$s of each period the laser pulse is applied 
(effective kicking strength for the lower hyperfine level of the ground state is $k\simeq6.5\times 10^{-4}$). We choose a small magnetic field $B=40\,$fT in $y$-direction so that we are well within the SERF regime, $R_\textrm{se}\gg\Omega_\textrm{Lar}$. 

With a circular polarized pump beam in $z$-direction resonant 
with the D1 line the initial spin-state is polarized which in 
the presence of spin-relaxation leads to an effective thermal state
\begin{equation}
\rho=\frac{e^{\beta K_z}e^{\beta S_z}}{Z_K Z_S},
\end{equation}
with the partition sum $Z_j=\sum_{m=-j}^je^{\beta m}$ and
$\beta=\ln\frac{1+q}{1-q}$, 
with polarization $q=0.95$. The readout is accomplished typically  
with the help of an off-resonant probe beam by measuring its 
polarization after it experienced a Faraday rotation when interacting 
with the atomic spin ensemble.
\section*{Supplementary References}


\section*{References}
\begin{thebibliography}{10}
	\expandafter\ifx\csname url\endcsname\relax
	\def\url#1{\texttt{#1}}\fi
	\expandafter\ifx\csname urlprefix\endcsname\relax\def\urlprefix{URL }\fi
	\providecommand{\bibinfo}[2]{#2}
	\providecommand{\eprint}[2][]{\url{#2}}
	
	\bibitem{Huelga97}
	\bibinfo{author}{Huelga, S.~F.} \emph{et~al.}
	\newblock \bibinfo{title}{Improvement of {F}requency {S}tandards with {Q}uantum
		{E}ntanglement}.
	\newblock \emph{\bibinfo{journal}{Phys. Rev. Lett.}}
	\textbf{\bibinfo{volume}{79}}, \bibinfo{pages}{3865--3868}
	(\bibinfo{year}{1997}).
	
	\bibitem{PhysRevLett.86.5870}
	\bibinfo{author}{Meyer, V.} \emph{et~al.}
	\newblock \bibinfo{title}{{Experimental Demonstration of Entanglement-Enhanced
			Rotation Angle Estimation Using Trapped Ions}}.
	\newblock \emph{\bibinfo{journal}{Phys. Rev. Lett.}}
	\textbf{\bibinfo{volume}{86}}, \bibinfo{pages}{5870--5873}
	(\bibinfo{year}{2001}).
	
	\bibitem{Leibfried04}
	\bibinfo{author}{Leibfried, D.} \emph{et~al.}
	\newblock \bibinfo{title}{Toward {H}eisenberg-{L}imited {S}pectroscopy with
		{M}ultiparticle {E}ntangled {S}tates}.
	\newblock \emph{\bibinfo{journal}{Science}} \textbf{\bibinfo{volume}{304}},
	\bibinfo{pages}{1476--1478} (\bibinfo{year}{2004}).
	
	\bibitem{wasilewski2010quantum}
	\bibinfo{author}{Wasilewski, W.} \emph{et~al.}
	\newblock \bibinfo{title}{Quantum {N}oise {L}imited and
		{E}ntanglement-{A}ssisted {M}agnetometry}.
	\newblock \emph{\bibinfo{journal}{Phys. Rev. Lett.}}
	\textbf{\bibinfo{volume}{104}}, \bibinfo{pages}{133601}
	(\bibinfo{year}{2010}).
	
	\bibitem{koschorreck2010sub}
	\bibinfo{author}{Koschorreck, M.}, \bibinfo{author}{Napolitano, M.},
	\bibinfo{author}{Dubost, B.} \& \bibinfo{author}{Mitchell, M.~W.}
	\newblock \bibinfo{title}{Sub-{P}rojection-{N}oise {S}ensitivity in {B}roadband
		{A}tomic {M}agnetometry}.
	\newblock \emph{\bibinfo{journal}{Phys. Rev. Lett.}}
	\textbf{\bibinfo{volume}{104}}, \bibinfo{pages}{093602}
	(\bibinfo{year}{2010}).
	
	\bibitem{Goda08}
	\bibinfo{author}{Goda, K.} \emph{et~al.}
	\newblock \bibinfo{title}{A quantum-enhanced prototype gravitational-wave
		detector}.
	\newblock \emph{\bibinfo{journal}{Nature Physics}}
	\textbf{\bibinfo{volume}{4}}, \bibinfo{pages}{472--476}
	(\bibinfo{year}{2008}).
	
	\bibitem{aasi2013enhanced}
	\bibinfo{author}{Aasi, J.} \emph{et~al.}
	\newblock \bibinfo{title}{Enhanced sensitivity of the {LIGO} gravitational wave
		detector by using squeezed states of light}.
	\newblock \emph{\bibinfo{journal}{Nature Photonics}}
	\textbf{\bibinfo{volume}{7}}, \bibinfo{pages}{613--619}
	(\bibinfo{year}{2013}).
	
	\bibitem{Giovannetti01}
	\bibinfo{author}{Giovannetti, V.}, \bibinfo{author}{Lloyd, S.} \&
	\bibinfo{author}{Maccone, L.}
	\newblock \bibinfo{title}{Quantum-enhanced positioning and clock
		synchronization}.
	\newblock \emph{\bibinfo{journal}{Nature}} \textbf{\bibinfo{volume}{412}},
	\bibinfo{pages}{417--419} (\bibinfo{year}{2001}).
	
	\bibitem{Allen08radar}
	\bibinfo{author}{Allen, E.~H.} \& \bibinfo{author}{Karageorgis, M.}
	\newblock \bibinfo{title}{Radar systems and methods using entangled quantum
		particles} (\bibinfo{year}{2008}).
	\newblock \bibinfo{note}{{US Patent 7,375,802}}.
	
	\bibitem{Taylor08}
	\bibinfo{author}{Taylor, J.} \emph{et~al.}
	\newblock \bibinfo{title}{High-sensitivity diamond magnetometer with nanoscale
		resolution}.
	\newblock \emph{\bibinfo{journal}{Nature Physics}}
	\textbf{\bibinfo{volume}{4}}, \bibinfo{pages}{810--816}
	(\bibinfo{year}{2008}).
	
	\bibitem{Boto00}
	\bibinfo{author}{Boto, A.~N.} \emph{et~al.}
	\newblock \bibinfo{title}{Quantum {I}nterferometric {O}ptical {L}ithography:
		{E}xploiting {E}ntanglement to {B}eat the {D}iffraction {L}imit}.
	\newblock \emph{\bibinfo{journal}{Phys. Rev. Lett.}}
	\textbf{\bibinfo{volume}{85}}, \bibinfo{pages}{2733--2736}
	(\bibinfo{year}{2000}).
	
	\bibitem{Higgins07}
	\bibinfo{author}{Higgins, B.~L.}, \bibinfo{author}{Berry, D.~W.},
	\bibinfo{author}{Bartlett, S.~D.}, \bibinfo{author}{Wiseman, H.~M.} \&
	\bibinfo{author}{Pryde, G.~J.}
	\newblock \bibinfo{title}{Entanglement-free {H}eisenberg-limited phase
		estimation}.
	\newblock \emph{\bibinfo{journal}{Nature}} \textbf{\bibinfo{volume}{450}},
	\bibinfo{pages}{393--396} (\bibinfo{year}{2007}).
	
	\bibitem{Nagata07}
	\bibinfo{author}{Nagata, T.}, \bibinfo{author}{Okamoto, R.},
	\bibinfo{author}{O'Brien, J.~L.} \& \bibinfo{author}{Takeuchi, K. S.~S.}
	\newblock \bibinfo{title}{Beating the {S}tandard {Q}uantum {L}imit with
		{F}our-{E}ntangled {P}hotons}.
	\newblock \emph{\bibinfo{journal}{Science}} \textbf{\bibinfo{volume}{316}},
	\bibinfo{pages}{726--729} (\bibinfo{year}{2007}).
	
	\bibitem{pezze_non-classical_2016}
	\bibinfo{author}{Pezz\`e, L.}, \bibinfo{author}{Smerzi, A.},
	\bibinfo{author}{Oberthaler, M.~K.}, \bibinfo{author}{Schmied, R.} \&
	\bibinfo{author}{Treutlein, P.}
	\newblock \bibinfo{title}{Quantum metrology with nonclassical states of atomic
		ensembles}.
	\newblock \eprint{Preprint at https://arxiv.org/abs/1609.01609v2 (2016)}.
	
	\bibitem{Peres91}
	\bibinfo{author}{Peres, A.}
	\newblock In \bibinfo{editor}{Cerdeira, H.~A.}, \bibinfo{editor}{Ramaswamy,
		R.}, \bibinfo{editor}{Gutzwiller, M.~C.} \& \bibinfo{editor}{Casati, G.}
	(eds.) \emph{\bibinfo{booktitle}{Quantum Chaos}} (\bibinfo{publisher}{World
		Scientific}, \bibinfo{address}{Singapore}, \bibinfo{year}{1991}).
	
	\bibitem{Gorin06review}
	\bibinfo{author}{Gorin, T.}, \bibinfo{author}{Prosen, T.},
	\bibinfo{author}{Seligman, T.~H.} \& \bibinfo{author}{{\v{Z}}nidari{\v{c}},
		M.}
	\newblock \bibinfo{title}{Dynamics of {L}oschmidt echoes and fidelity decay}.
	\newblock \emph{\bibinfo{journal}{Physics Reports}}
	\textbf{\bibinfo{volume}{435}}, \bibinfo{pages}{33--156}
	(\bibinfo{year}{2006}).
	
	\bibitem{Boixo08.2}
	\bibinfo{author}{Boixo, S.} \emph{et~al.}
	\newblock \bibinfo{title}{Quantum {M}etrology: Dynamics versus {E}ntanglement}.
	\newblock \emph{\bibinfo{journal}{Phys. Rev. Lett.}}
	\textbf{\bibinfo{volume}{101}}, \bibinfo{pages}{040403}
	(\bibinfo{year}{2008}).
	
	\bibitem{xiao2011chaos}
	\bibinfo{author}{Xiao-Qian, W.}, \bibinfo{author}{Jian, M.},
	\bibinfo{author}{Xi-He, Z.} \& \bibinfo{author}{Xiao-Guang, W.}
	\newblock \bibinfo{title}{Chaos and quantum {F}isher information in the quantum
		kicked top}.
	\newblock \emph{\bibinfo{journal}{Chinese Physics B}}
	\textbf{\bibinfo{volume}{20}}, \bibinfo{pages}{050510}
	(\bibinfo{year}{2011}).
	
	\bibitem{song2012quantum}
	\bibinfo{author}{Song, L.}, \bibinfo{author}{Ma, J.}, \bibinfo{author}{Yan, D.}
	\& \bibinfo{author}{Wang, X.}
	\newblock \bibinfo{title}{Quantum {F}isher information and chaos in the {D}icke
		model}.
	\newblock \emph{\bibinfo{journal}{The European Physical Journal D}}
	\textbf{\bibinfo{volume}{66}}, \bibinfo{pages}{201} (\bibinfo{year}{2012}).
	
	\bibitem{weiss2009signatures}
	\bibinfo{author}{Weiss, C.} \& \bibinfo{author}{Teichmann, N.}
	\newblock \bibinfo{title}{Signatures of chaos-induced mesoscopic entanglement}.
	\newblock \emph{\bibinfo{journal}{Journal of Physics B: Atomic, Molecular and
			Optical Physics}} \textbf{\bibinfo{volume}{42}}, \bibinfo{pages}{031001}
	(\bibinfo{year}{2009}).
	
	\bibitem{frowis2016detecting}
	\bibinfo{author}{Fr{\"o}wis, F.}, \bibinfo{author}{Sekatski, P.} \&
	\bibinfo{author}{D{\"u}r, W.}
	\newblock \bibinfo{title}{{Detecting Large Quantum Fisher Information with
			Finite Measurement Precision}}.
	\newblock \emph{\bibinfo{journal}{Phys. Rev. Lett.}}
	\textbf{\bibinfo{volume}{116}}, \bibinfo{pages}{090801}
	(\bibinfo{year}{2016}).
	
	\bibitem{macri2016loschmidt}
	\bibinfo{author}{Macr{\`\i}, T.}, \bibinfo{author}{Smerzi, A.} \&
	\bibinfo{author}{Pezz{\`e}, L.}
	\newblock \bibinfo{title}{Loschmidt echo for quantum metrology}.
	\newblock \emph{\bibinfo{journal}{Phys. Rev. A}} \textbf{\bibinfo{volume}{94}},
	\bibinfo{pages}{010102} (\bibinfo{year}{2016}).
	
	\bibitem{linnemann2016quantum}
	\bibinfo{author}{Linnemann, D.} \emph{et~al.}
	\newblock \bibinfo{title}{Quantum-{E}nhanced {S}ensing {B}ased on {T}ime
		{R}eversal of {N}onlinear {D}ynamics}.
	\newblock \emph{\bibinfo{journal}{Phys. Rev. Lett.}}
	\textbf{\bibinfo{volume}{117}}, \bibinfo{pages}{013001}
	(\bibinfo{year}{2016}).
	
	\bibitem{ma2011quantum}
	\bibinfo{author}{Ma, J.}, \bibinfo{author}{Wang, X.}, \bibinfo{author}{Sun, C.}
	\& \bibinfo{author}{Nori, F.}
	\newblock \bibinfo{title}{Quantum spin squeezing}.
	\newblock \emph{\bibinfo{journal}{Physics Reports}}
	\textbf{\bibinfo{volume}{509}}, \bibinfo{pages}{89--165}
	(\bibinfo{year}{2011}).
	
	\bibitem{madhok2014information}
	\bibinfo{author}{Madhok, V.}, \bibinfo{author}{Riofr{\'\i}o, C.~A.},
	\bibinfo{author}{Ghose, S.} \& \bibinfo{author}{Deutsch, I.~H.}
	\newblock \bibinfo{title}{Information {G}ain in {T}omography--{A} {Q}uantum
		{S}ignature of {C}haos}.
	\newblock \emph{\bibinfo{journal}{Phys. Rev. Lett.}}
	\textbf{\bibinfo{volume}{112}}, \bibinfo{pages}{014102}
	(\bibinfo{year}{2014}).
	
	\bibitem{madhok2016review}
	\bibinfo{author}{Madhok, V.}, \bibinfo{author}{Riofrio, C.~A.} \&
	\bibinfo{author}{Deutsch, I.~H.}
	\newblock \bibinfo{title}{Review: Characterizing and quantifying quantum chaos
		with quantum tomography}.
	\newblock \emph{\bibinfo{journal}{Pramana}} \textbf{\bibinfo{volume}{87}},
	\bibinfo{pages}{65} (\bibinfo{year}{2016}).
	
	\bibitem{braun2017without}
	\bibinfo{author}{Braun, D.} \emph{et~al.}
	\newblock \bibinfo{title}{Quantum enhanced measurements without entanglement}.
	\newblock \eprint{Preprint at https://arxiv.org/abs/1701.05152 (2017)}.
	
	\bibitem{allred2002high}
	\bibinfo{author}{Allred, J.}, \bibinfo{author}{Lyman, R.},
	\bibinfo{author}{Kornack, T.} \& \bibinfo{author}{Romalis, M.}
	\newblock \bibinfo{title}{{High-Sensitivity Atomic Magnetometer Unaffected by
			Spin-Exchange Relaxation}}.
	\newblock \emph{\bibinfo{journal}{Phys. Rev. Lett.}}
	\textbf{\bibinfo{volume}{89}}, \bibinfo{pages}{130801}
	(\bibinfo{year}{2002}).
	
	\bibitem{Kominis03}
	\bibinfo{author}{Kominis, I.~K.}, \bibinfo{author}{Kornack, T.~W.},
	\bibinfo{author}{Allred, J.~C.} \& \bibinfo{author}{Romalis, M.~V.}
	\newblock \bibinfo{title}{A subfemtotesla multichannel atomic magnetometer}.
	\newblock \emph{\bibinfo{journal}{Nature}} \textbf{\bibinfo{volume}{422}},
	\bibinfo{pages}{596–--599} (\bibinfo{year}{2003}).
	
	\bibitem{savukov2005effects}
	\bibinfo{author}{Savukov, I.} \& \bibinfo{author}{Romalis, M.}
	\newblock \bibinfo{title}{Effects of spin-exchange collisions in a high-density
		alkali-metal vapor in low magnetic fields}.
	\newblock \emph{\bibinfo{journal}{Phys. Rev. A}} \textbf{\bibinfo{volume}{71}},
	\bibinfo{pages}{023405} (\bibinfo{year}{2005}).
	
	\bibitem{budker2007optical}
	\bibinfo{author}{Budker, D.} \& \bibinfo{author}{Romalis, M.}
	\newblock \bibinfo{title}{Optical magnetometry}.
	\newblock \emph{\bibinfo{journal}{Nature Physics}}
	\textbf{\bibinfo{volume}{3}}, \bibinfo{pages}{227--234}
	(\bibinfo{year}{2007}).
	
	\bibitem{sheng2013subfemtotesla}
	\bibinfo{author}{Sheng, D.}, \bibinfo{author}{Li, S.}, \bibinfo{author}{Dural,
		N.} \& \bibinfo{author}{Romalis, M.}
	\newblock \bibinfo{title}{{Subfemtotesla Scalar Atomic Magnetometry Using
			Multipass Cells}}.
	\newblock \emph{\bibinfo{journal}{Phys. Rev. Lett.}}
	\textbf{\bibinfo{volume}{110}}, \bibinfo{pages}{160802}
	(\bibinfo{year}{2013}).
	
	\bibitem{appelt1998theory}
	\bibinfo{author}{Appelt, S.} \emph{et~al.}
	\newblock \bibinfo{title}{Theory of spin-exchange optical pumping of $^3${H}e
		and $^{129}${X}e}.
	\newblock \emph{\bibinfo{journal}{Phys. Rev. A}} \textbf{\bibinfo{volume}{58}},
	\bibinfo{pages}{1412--1439} (\bibinfo{year}{1998}).
	
	\bibitem{haake1987classical}
	\bibinfo{author}{Haake, F.}, \bibinfo{author}{Ku{\'s}, M.} \&
	\bibinfo{author}{Scharf, R.}
	\newblock \bibinfo{title}{Classical and quantum chaos for a kicked top}.
	\newblock \emph{\bibinfo{journal}{Zeitschrift f{\"u}r Physik B Condensed
			Matter}} \textbf{\bibinfo{volume}{65}}, \bibinfo{pages}{381--395}
	(\bibinfo{year}{1987}).
	
	\bibitem{kickedtop}
	\bibinfo{author}{Haake, F.}, \bibinfo{author}{Ku\'s, M.} \&
	\bibinfo{author}{Scharf, R.}
	\newblock In \bibinfo{editor}{Haake, F.}, \bibinfo{editor}{Narducci, L.} \&
	\bibinfo{editor}{Walls, D.} (eds.) \emph{\bibinfo{booktitle}{Coherence,
			Cooperation, and Fluctuations}} (\bibinfo{publisher}{Cambridge University
		Press}, \bibinfo{address}{Cambridge}, \bibinfo{year}{1986}).
	
	\bibitem{haake2013quantum}
	\bibinfo{author}{Haake, F.}
	\newblock \emph{\bibinfo{title}{Quantum {S}ignatures of {C}haos}},
	vol.~\bibinfo{volume}{54} (\bibinfo{publisher}{Springer Science \& Business
		Media}, \bibinfo{address}{Berlin}, \bibinfo{year}{2013}).
	
	\bibitem{haake2000can}
	\bibinfo{author}{Haake, F.}
	\newblock \bibinfo{title}{Can the kicked top be realized?}
	\newblock \emph{\bibinfo{journal}{Journal of Modern Optics}}
	\textbf{\bibinfo{volume}{47}}, \bibinfo{pages}{2883--2890}
	(\bibinfo{year}{2000}).
	
	\bibitem{chaudhury2009quantum}
	\bibinfo{author}{Chaudhury, S.}, \bibinfo{author}{Smith, A.},
	\bibinfo{author}{Anderson, B.}, \bibinfo{author}{Ghose, S.} \&
	\bibinfo{author}{Jessen, P.~S.}
	\newblock \bibinfo{title}{Quantum signatures of chaos in a kicked top}.
	\newblock \emph{\bibinfo{journal}{Nature}} \textbf{\bibinfo{volume}{461}},
	\bibinfo{pages}{768--771} (\bibinfo{year}{2009}).
	
	\bibitem{Helstrom1969}
	\bibinfo{author}{Helstrom, C.~W.}
	\newblock \bibinfo{title}{Quantum detection and estimation theory}.
	\newblock \emph{\bibinfo{journal}{Journal of Statistical Physics}}
	\textbf{\bibinfo{volume}{1}}, \bibinfo{pages}{231--252}
	(\bibinfo{year}{1969}).
	
	\bibitem{Miszczak09}
	\bibinfo{author}{Miszczak, J.~A.}, \bibinfo{author}{Pucha\l{}a, Z.},
	\bibinfo{author}{Horodecki, P.}, \bibinfo{author}{Uhlmann, A.} \&
	\bibinfo{author}{K.\.{Z}yczkowski}.
	\newblock \bibinfo{title}{Sub- and super-fidelity as bounds for quantum
		fidelity}.
	\newblock \emph{\bibinfo{journal}{Quantum Information and Computation}}
	\textbf{\bibinfo{volume}{9}}, \bibinfo{pages}{0103--0130}
	(\bibinfo{year}{2009}).
	
	\bibitem{Braunstein94}
	\bibinfo{author}{Braunstein, S.~L.} \& \bibinfo{author}{Caves, C.~M.}
	\newblock \bibinfo{title}{Statistical distance and the geometry of quantum
		states}.
	\newblock \emph{\bibinfo{journal}{Phys. Rev. Lett.}}
	\textbf{\bibinfo{volume}{72}}, \bibinfo{pages}{3439--3443}
	(\bibinfo{year}{1994}).
	
	\bibitem{Braunstein90}
	\bibinfo{author}{Braunstein, S.~L.} \& \bibinfo{author}{Caves, C.~M.}
	\newblock \bibinfo{title}{Wringing out better {B}ell inequalities}.
	\newblock \emph{\bibinfo{journal}{Annals of Physics}}
	\textbf{\bibinfo{volume}{202}}, \bibinfo{pages}{22--56}
	(\bibinfo{year}{1990}).
	
	\bibitem{zaslavsky1981stochasticity}
	\bibinfo{author}{Zaslavsky, G.~M.}
	\newblock \bibinfo{title}{Stochasticity in quantum systems}.
	\newblock \emph{\bibinfo{journal}{Physics Reports}}
	\textbf{\bibinfo{volume}{80}}, \bibinfo{pages}{157--250}
	(\bibinfo{year}{1981}).
	
	\bibitem{PhysRevE.64.055203}
	\bibinfo{author}{Jacquod, P.}, \bibinfo{author}{Silvestrov, P.} \&
	\bibinfo{author}{Beenakker, C.}
	\newblock \bibinfo{title}{Golden rule decay versus {Lyapunov} decay of the
		quantum {Loschmidt} echo}.
	\newblock \emph{\bibinfo{journal}{Phys. Rev. E}} \textbf{\bibinfo{volume}{64}},
	\bibinfo{pages}{055203} (\bibinfo{year}{2001}).
	
	\bibitem{PhysRevE.65.066205}
	\bibinfo{author}{Benenti, G.} \& \bibinfo{author}{Casati, G.}
	\newblock \bibinfo{title}{Quantum-classical correspondence in perturbed chaotic
		systems}.
	\newblock \emph{\bibinfo{journal}{Phys. Rev. E}} \textbf{\bibinfo{volume}{65}},
	\bibinfo{pages}{066205} (\bibinfo{year}{2002}).
	
	\bibitem{sankaranarayanan2003recurrence}
	\bibinfo{author}{Sankaranarayanan, R.} \& \bibinfo{author}{Lakshminarayan, A.}
	\newblock \bibinfo{title}{Recurrence of fidelity in nearly integrable systems}.
	\newblock \emph{\bibinfo{journal}{Phys. Rev. E}} \textbf{\bibinfo{volume}{68}},
	\bibinfo{pages}{036216} (\bibinfo{year}{2003}).
	
	\bibitem{Dicke54}
	\bibinfo{author}{Dicke, R.~H.}
	\newblock \bibinfo{title}{Coherence in {S}pontaneous {R}adiation {P}rocesses}.
	\newblock \emph{\bibinfo{journal}{Phys. Rev.}} \textbf{\bibinfo{volume}{93}},
	\bibinfo{pages}{99--110} (\bibinfo{year}{1954}).
	
	\bibitem{Bonifacio71a}
	\bibinfo{author}{Bonifacio, R.}, \bibinfo{author}{Schwendiman, P.} \&
	\bibinfo{author}{Haake, F.}
	\newblock \bibinfo{title}{Quantum {S}tatistical {T}heory of {S}uperradiance
		{I}}.
	\newblock \emph{\bibinfo{journal}{Phys. Rev. A}} \textbf{\bibinfo{volume}{4}},
	\bibinfo{pages}{302--313} (\bibinfo{year}{1971}).
	
	\bibitem{Haake73}
	\bibinfo{author}{Haake, F.}
	\newblock \emph{\bibinfo{title}{{Statistical {T}reatment of {O}pen {S}ystems by
				{G}eneralized {M}aster {E}quations}}}, vol.~\bibinfo{volume}{66} of
	\emph{\bibinfo{series}{Springer Tracts in Modern Physics}}
	(\bibinfo{publisher}{Springer}, \bibinfo{address}{Berlin},
	\bibinfo{year}{1973}).
	
	\bibitem{Gross82}
	\bibinfo{author}{Gross, M.} \& \bibinfo{author}{Haroche, S.}
	\newblock \bibinfo{title}{Superradiance: An {E}ssay on the {T}heory of
		{C}ollective {S}pontaneous {E}mission}.
	\newblock \emph{\bibinfo{journal}{Phys. Rep.}} \textbf{\bibinfo{volume}{93}},
	\bibinfo{pages}{301--396} (\bibinfo{year}{1982}).
	
	\bibitem{kaluzny1983observation}
	\bibinfo{author}{Kaluzny, Y.}, \bibinfo{author}{Goy, P.},
	\bibinfo{author}{Gross, M.}, \bibinfo{author}{Raimond, J.} \&
	\bibinfo{author}{Haroche, S.}
	\newblock \bibinfo{title}{{Observation of Self-Induced Rabi Oscillations in
			Two-Level Atoms Excited Inside a Resonant Cavity: The Ringing Regime of
			Superradiance}}.
	\newblock \emph{\bibinfo{journal}{Phys. Rev. Lett.}}
	\textbf{\bibinfo{volume}{51}}, \bibinfo{pages}{1175--1178}
	(\bibinfo{year}{1983}).
	
	\bibitem{Braun01B}
	\bibinfo{author}{Braun, D.}
	\newblock \emph{\bibinfo{title}{Dissipative Quantum Chaos and Decoherence}},
	vol. \bibinfo{volume}{172} of \emph{\bibinfo{series}{Springer Tracts in
			Modern Physics}} (\bibinfo{publisher}{Springer}, \bibinfo{address}{Berlin},
	\bibinfo{year}{2001}).
	
	\bibitem{balabas2010polarized}
	\bibinfo{author}{Balabas, M.}, \bibinfo{author}{Karaulanov, T.},
	\bibinfo{author}{Ledbetter, M.} \& \bibinfo{author}{Budker, D.}
	\newblock \bibinfo{title}{{Polarized Alkali-Metal Vapor with Minute-Long
			Transverse Spin-Relaxation Time}}.
	\newblock \emph{\bibinfo{journal}{Phys. Rev. Lett.}}
	\textbf{\bibinfo{volume}{105}}, \bibinfo{pages}{070801}
	(\bibinfo{year}{2010}).
	
	\bibitem{deutsch2010quantum}
	\bibinfo{author}{Deutsch, I.~H.} \& \bibinfo{author}{Jessen, P.~S.}
	\newblock \bibinfo{title}{Quantum control and measurement of atomic spins in
		polarization spectroscopy}.
	\newblock \emph{\bibinfo{journal}{Optics Communications}}
	\textbf{\bibinfo{volume}{283}}, \bibinfo{pages}{681--694}
	(\bibinfo{year}{2010}).
	
	\bibitem{takeuchi2005spin}
	\bibinfo{author}{Takeuchi, M.} \emph{et~al.}
	\newblock \bibinfo{title}{{Spin Squeezing via One-Axis Twisting with Coherent
			Light}}.
	\newblock \emph{\bibinfo{journal}{Phys. Rev. Lett.}}
	\textbf{\bibinfo{volume}{94}}, \bibinfo{pages}{023003}
	(\bibinfo{year}{2005}).
	
	\bibitem{ledbetter2008spin}
	\bibinfo{author}{Ledbetter, M.}, \bibinfo{author}{Savukov, I.},
	\bibinfo{author}{Acosta, V.}, \bibinfo{author}{Budker, D.} \&
	\bibinfo{author}{Romalis, M.}
	\newblock \bibinfo{title}{Spin-exchange-relaxation-free magnetometry with {C}s
		vapor}.
	\newblock \emph{\bibinfo{journal}{Phys. Rev. A}} \textbf{\bibinfo{volume}{77}},
	\bibinfo{pages}{033408} (\bibinfo{year}{2008}).
	
	\bibitem{agarwal1997atomic}
	\bibinfo{author}{Agarwal, G.}, \bibinfo{author}{Puri, R.} \&
	\bibinfo{author}{Singh, R.}
	\newblock \bibinfo{title}{Atomic {S}chr{\"o}dinger cat states}.
	\newblock \emph{\bibinfo{journal}{Phys. Rev. A}} \textbf{\bibinfo{volume}{56}},
	\bibinfo{pages}{2249--2254} (\bibinfo{year}{1997}).
	
	\bibitem{Julsgaard01}
	\bibinfo{author}{Julsgaard, B.}, \bibinfo{author}{Kozhekin, A.} \&
	\bibinfo{author}{Polzik, E.~S.}
	\newblock \bibinfo{title}{Experimental long-lived entanglement of two
		macroscopic objects}.
	\newblock \emph{\bibinfo{journal}{Nature}} \textbf{\bibinfo{volume}{413}},
	\bibinfo{pages}{400--403} (\bibinfo{year}{2001}).
	
\end{thebibliography}

\begin{thebibliography}{1}
	\expandafter\ifx\csname url\endcsname\relax
	\def\url#1{\texttt{#1}}\fi
	\expandafter\ifx\csname urlprefix\endcsname\relax\def\urlprefix{URL }\fi
	\providecommand{\bibinfo}[2]{#2}
	\providecommand{\eprint}[2][]{\url{#2}}
	
	\bibitem{chaudhury2009quantumsup}
	\bibinfo{author}{Chaudhury, S.}, \bibinfo{author}{Smith, A.},
	\bibinfo{author}{Anderson, B.}, \bibinfo{author}{Ghose, S.} \&
	\bibinfo{author}{Jessen, P.~S.}
	\newblock \bibinfo{title}{Quantum signatures of chaos in a kicked top}.
	\newblock \emph{\bibinfo{journal}{Nature}} \textbf{\bibinfo{volume}{461}},
	\bibinfo{pages}{768--771} (\bibinfo{year}{2009}).
	
	\bibitem{haake2000cansup}
	\bibinfo{author}{Haake, F.}
	\newblock \bibinfo{title}{Can the kicked top be realized?}
	\newblock \emph{\bibinfo{journal}{Journal of Modern Optics}}
	\textbf{\bibinfo{volume}{47}}, \bibinfo{pages}{2883--2890}
	(\bibinfo{year}{2000}).
	
	\bibitem{agarwal1997atomicsup}
	\bibinfo{author}{Agarwal, G.}, \bibinfo{author}{Puri, R.} \&
	\bibinfo{author}{Singh, R.}
	\newblock \bibinfo{title}{Atomic {S}chr{\"o}dinger cat states}.
	\newblock \emph{\bibinfo{journal}{Phys. Rev. A}} \textbf{\bibinfo{volume}{56}},
	\bibinfo{pages}{2249--2254} (\bibinfo{year}{1997}).
	
	\bibitem{balabas2010polarizedsup}
	\bibinfo{author}{Balabas, M.}, \bibinfo{author}{Karaulanov, T.},
	\bibinfo{author}{Ledbetter, M.} \& \bibinfo{author}{Budker, D.}
	\newblock \bibinfo{title}{{Polarized Alkali-Metal Vapor with Minute-Long
			Transverse Spin-Relaxation Time}}.
	\newblock \emph{\bibinfo{journal}{Phys. Rev. Lett.}}
	\textbf{\bibinfo{volume}{105}}, \bibinfo{pages}{070801}
	(\bibinfo{year}{2010}).
	
	\bibitem{appelt1998theorysup}
	\bibinfo{author}{Appelt, S.} \emph{et~al.}
	\newblock \bibinfo{title}{Theory of spin-exchange optical pumping of $^3${H}e
		and $^{129}${X}e}.
	\newblock \emph{\bibinfo{journal}{Phys. Rev. A}} \textbf{\bibinfo{volume}{58}},
	\bibinfo{pages}{1412--1439} (\bibinfo{year}{1998}).
	
	\bibitem{deutsch2010quantumsup}
	\bibinfo{author}{Deutsch, I.~H.} \& \bibinfo{author}{Jessen, P.~S.}
	\newblock \bibinfo{title}{Quantum control and measurement of atomic spins in
		polarization spectroscopy}.
	\newblock \emph{\bibinfo{journal}{Optics Communications}}
	\textbf{\bibinfo{volume}{283}}, \bibinfo{pages}{681--694}
	(\bibinfo{year}{2010}).
	
	\bibitem{savukov2005effectssup}
	\bibinfo{author}{Savukov, I.} \& \bibinfo{author}{Romalis, M.}
	\newblock \bibinfo{title}{Effects of spin-exchange collisions in a high-density
		alkali-metal vapor in low magnetic fields}.
	\newblock \emph{\bibinfo{journal}{Phys. Rev. A}} \textbf{\bibinfo{volume}{71}},
	\bibinfo{pages}{023405} (\bibinfo{year}{2005}).
	
\end{thebibliography}
\end{document}